\documentclass[fleqn,10pt]{wlscirep}
\usepackage[utf8]{inputenc}
\usepackage[T1]{fontenc}

\usepackage{amsmath}
\usepackage{amsfonts}
\usepackage{graphicx}
\usepackage{amsthm}
\usepackage{moreverb}
\usepackage{amsfonts}
\usepackage{mathrsfs}
\usepackage{float}
\usepackage{gensymb}
\usepackage{xfrac}
\usepackage{color}
\usepackage{caption}
\usepackage{subcaption}
\usepackage{epstopdf}
\usepackage{listings}
\usepackage{todonotes}
\usepackage{booktabs}
\usepackage{array}
\usepackage{tabularx}
\usepackage{placeins}

\makeatletter
\newcommand{\thickhline}{%
    \noalign {\ifnum 0=`}\fi \hrule height 1pt
    \futurelet \reserved@a \@xhline
}
\newcolumntype{"}{@{\hskip\tabcolsep\vrule width 1pt\hskip\tabcolsep}}
\makeatother

\usepackage{array,multirow,graphicx}
\usepackage[ruled, lined, linesnumbered, commentsnumbered, longend]{algorithm2e}

\SetKwInput{KwInput}{Input}
\SetKwInput{KwOutput}{Output}

\title{The Deep Latent Space Particle Filter for Real-Time Data Assimilation with Uncertainty Quantification}

\author[1, 3 , *]{Nikolaj T. M{\"u}cke}
\author[2, 4]{Sander M. Bohté}
\author[3]{Cornelis W. Oosterlee}

\affil[1]{Centrum Wiskunde \& Informatica, Scientific Computing, Amsterdam, 1098 XG, The Netherlands}
\affil[2]{Centrum Wiskunde \& Informatica, Machine Learning, Amsterdam, 1098 XG, The Netherlands}
\affil[3]{Utrecht University, Mathematical Institute, Utrecht,  3584 CS, The Netherlands}
\affil[4]{University of Amsterdam, Swammerdam Institute of Life Sciences (SILS), Amsterdam, 1098 XH, The Netherlands}

\affil[*]{nikolaj.mucke@cwi.nl}

\keywords{Particle filter, transformers, Wasserstein autoencoders, partial differential equations, data assimilation}

\begin{abstract}
In Data Assimilation, observations are fused with simulations to obtain an accurate estimate of the state and parameters for a given physical system. Combining data with a model, however, while accurately estimating uncertainty, is computationally expensive and infeasible to run in real-time for complex systems. Here, we present a novel particle filter methodology, the Deep Latent Space Particle filter or {\em D-LSPF},  that uses neural network-based surrogate models to overcome this computational challenge. The D-LSPF enables filtering in the low-dimensional latent space obtained using Wasserstein AEs with modified vision transformer layers for dimensionality reduction and transformers for parameterized latent space time stepping. As we demonstrate on three test cases, including leak localization in multi-phase pipe flow and seabed identification for fully nonlinear water waves, the D-LSPF runs orders of magnitude faster than a high-fidelity particle filter and 3-5 times faster than alternative methods while being up to an order of magnitude more accurate. The D-LSPF thus enables real-time data assimilation with uncertainty quantification for physical systems.
\end{abstract}
\begin{document}

\flushbottom
\maketitle
%
%
\thispagestyle{empty}

\section*{Introduction} \label{paper_4:section:introduction}

Virtual representations of physical systems like digital twins have proven to be invaluable tools for monitoring, predicting, and optimizing the performance of intricate systems, ranging from industrial machinery to biological processes \cite{rasheed2020digital}. The efficacy of digital twins relies however on the accurate assimilation of real-time data into the simulations to ensure accurate calibration and state estimation in situations where the state and its dynamics are not known. Importantly, to confidently rely on the information provided, data assimilation should be accompanied by a quantification of the associated uncertainty originating from both measurements and model errors \cite{asch2022toolbox, kapteyn2021probabilistic}.

Performing data assimilation with uncertainty quantification in real time for high-dimensional systems, such as discretized partial differential equations (PDEs), is computationally infeasible due to the need to compute large ensembles of solutions. Approaches such as (ensemble) Kalman filtering \cite{houtekamer1998data} aim to overcome this computational bottleneck by assuming Gaussian distributed prior, likelihood, and posterior distributions \cite{evensen2022data}, which is restrictive in practical situations where these assumptions do not hold \cite{uilhoorn2014particle, albarakati2022model}. Particle filters, on the other hand, can approximate any distribution provided there is a sufficient number of particles in the ensemble \cite{kitagawa1996monte, fearnhead2018particle}. The necessary ensemble size for particle filters however often makes it infeasible to run in real-time for complex, high-dimensional systems. Therefore, there is a need for methods to speed up the computations of ensembles. 

To speed up the computation of ensembles, surrogate models are used to approximate the forward problem by replacing the full order model with a computationally cheaper alternative. Surrogate models based on proper orthogonal decomposition and dynamic mode decomposition have been developed with reasonable success \cite{albarakati2022model}. However, for nonlinear, hyperbolic, and/or discontinuous problems, advanced surrogates are necessary to achieve the desired speed-up \cite{lee2020model}. Therefore, deep learning approaches have received increased attention in efforts to obtain significant speed-ups without sacrificing essential accuracy \cite{lee2020model, geneva2022transformers, champion2019data, mucke2021reduced}. 

Deep learning-based surrogate models can be designed in various ways. One approach is to make use of latent space representations. Here, high-dimensional states are mapped onto a low-dimensional latent space such that the expensive computations, such as time stepping, can be performed cheaply in this latent space. The autoencoder (AE) \cite{ballard1987modular} is the principal enabling architecture for this. In the AE, an encoder network reduces the original data into a latent representation and a decoder subsequently reconstructs the original data from the latent representation. Since \cite{ballard1987modular}, many extensions and improvements have been developed, such as adding probabilistic priors to the latent space \cite{kingma2013auto, tolstikhin2017wasserstein}. In the context of surrogate modeling for physical systems, the focus has been on ensuring that AEs learn latent representations that are suitable for downstream tasks, such as time stepping, via various regularization techniques \cite{champion2019data, geneva2022transformers, wan2023evolve}. The success of such regularization is important when embedding the surrogate model into a data assimilation framework.

Utilizing neural network surrogate models for speeding up data assimilation has been explored in various studies, see also \cite{cheng2023machine} for a review. In \cite{patel2020bayesian, xia2022bayesian, mucke2023markov, seabra2024ai}, generative deep learning has been used for high-dimensional state- and parameter estimation. However, none of these approaches performed sequential assimilation of the data. In \cite{chen2023reduced, brunton2016discovering}, deep learning was used for model discovery. In such applications, the model is a-priori unknown and is learned from observations, typically requiring a huge number of observations in space and time to recover the complete, unknown state; to this point it is not clear how these methods perform with real-time data assimilation. Similarly, in \cite{maddison2017filtering, moretti2019particle} a particle filter approach to data assimilation has been used to formulate a variational objective for training a latent space model. Hence, the latent model is trained while data is arriving, which severely limits  real-time assimilation for high-dimensional, nonlinear problems with limited observations. In \cite{silva2023generative}, a GAN set-up, combined with proper orthogonal decomposition, was used for sequential data assimilation with an approach similar to randomized maximum likelihood. However, unlike particle filters, convergence of such methods is not ensured for nonlinear cases \cite{evensen2022data}. 

Yet, while many deep learning-based surrogate models have been used to speed up data assimilation, there is limited work on such approaches using particle filters \cite{gonczarek2016articulated, yang2022particle}. In \cite{gonczarek2016articulated} a back-constrained Gaussian process latent variable model is used to parameterize both the dimensionality reduction and latent space dynamics. In \cite{yang2022particle}, a particle filter using a latent space formulation was presented. The approach evaluated the likelihood by iterative closest point registration fitness scores and the latent time stepping was mainly linear. Since we are only dealing with highly nonlinear PDE-based problems with very few observations in space and time in this paper, we have chosen not implement and compare with these methods, as it is unlikely that linear time stepping will be sufficient.


Here, we propose a deep learning framework for performing particle filtering in real-time using latent-space representations: the Deep Latent Space Particle Filter, or {\em D-LSPF}, targeting complex nonlinear data assimilation problems modeled by PDEs. For this, we develop a novel extension to the vision transformer layer for dimensionality reduction of the high-dimensional state in an AE setup. A transformer-based network is then used for parameterized time stepping, which enables filtering in the latent space as well parameter estimation. To ensure that the latent space has the appropriate desirable properties, we combine several regularization techniques such as divergence and consistency regularization \cite{tolstikhin2017wasserstein, wan2023evolve}. We showcase the D-LSPF on three distinct test problems with varying characteristics, such as discontinuity, few observations in space and time, parameter estimation, and highly oscillatory real-world data. In all cases, the D-LSPF demonstrates significant speed-ups compared to alternative methods without sacrificing accuracy. This promises to enable true or near real-time data assimilation for new, more complex classes of problems, with direct applications in engineering such as leak localization as well as seabed and wave height estimation. 

The paper is organized as follows. In the second section we describe the problem setting. This consists of a brief description of the Bayesian filtering problem and an overview of the particle filter. In the next section, we present the D-LSPF. Firstly, we outline the latent filtering problem, followed by a description of the latent space regularized AE using the novel transformer-based dimensionality reduction layers, and parameterized time stepping. Lastly, we showcase the performance of the D-LSPF on three test cases, namely the viscous Burgers equation, harmonic wave generation over a submerged bar, and leak localization for multi-phase flow in a pipe. The D-LSPF is compared with a high-fidelity particle filter and the Reduced-Order Autodifferentiable Ensemble Kalman Filter (ROAD-EnKF) \cite{chen2023reduced}.

\section*{Problem Setting} \label{paper_4:section:problem_statement}
We consider problems that are modeled by time-dependent PDEs. Such problems consist of a state, typically made up by several quantities such as velocity and pressure, and parameters, source terms, boundary conditions, and initial conditions. Since the model won't be perfect and true values of the parameters are rarely known, the problem needs to be accompanied by observations coming from a series of sensors. However, sensors deliver noisy data and are often scarcely placed in the domain of interest, so that the data needs to be assimilated into the model to yield an accurate estimate of the state and parameters. 

Consider a time and spatially discretized PDE, with accompanying observations: \vspace{-5pt}
\begin{align} \label{paper_4:eq:state_space_model}
\begin{aligned}[c]
    \boldsymbol{q}_{n} &= F(\boldsymbol{q}_{n-1};\boldsymbol{m}_{n-1}) + \boldsymbol{\xi}_{n-1}, \\
    \boldsymbol{m}_{n} &= G(\boldsymbol{m}_{n-1}) + \boldsymbol{\zeta}_{n-1}, \\
    \boldsymbol{y}_{n} &= h(\boldsymbol{q}_{n}) + \boldsymbol{\eta}_{n-1},
\end{aligned}
\quad
\begin{aligned}[c]
     &\boldsymbol{\xi}_{n-1}\sim P_\xi(\boldsymbol{\xi}_{n-1}), \\
     &\boldsymbol{\zeta}_{n-1}\sim P_\zeta(\boldsymbol{\zeta}_{n-1}), \\
     &\boldsymbol{\eta}_{n-1}\sim P_\eta(\boldsymbol{\eta}_{n-1}),
\end{aligned} 
\end{align} 
where $F$ is a (nonlinear) operator advancing the state, $G$ is an operator advancing the parameters, $\boldsymbol{q}_n(\boldsymbol{m}_n) \in \mathbb{R}^{N_x}$ is a parameter-dependent state at time step $n$, $\boldsymbol{m}_n\in \mathbb{R}^{N_m}$ are the parameters, $\boldsymbol{y}_n\in \mathbb{R}^{N_o}$ is an observation vector at time step $n$, $h: \mathbb{R}^{N_x}\rightarrow \mathbb{R}^{N_o}$ is the observation operator, $\boldsymbol{\xi}_n$ is the model error, $\boldsymbol{\zeta}_n$ is the parameter error, and $\boldsymbol{\eta}_n$ is the observation noise. Note that when parameters are constant in time, we use $G(\boldsymbol{m}_n)=\boldsymbol{m}_n$.
To simplify notation, we introduce the combined state-parameter variable, $\boldsymbol{u}_n = (\boldsymbol{q}_n, \boldsymbol{m}_n)$. We will refer to $\boldsymbol{u}_n$ as the augmented state and introduce the notation for time series, $\boldsymbol{u}_{0:n} = (\boldsymbol{u}_{0}, \ldots, \boldsymbol{u}_{n})$.
We further assume that the probability density functions exist and will therefore continue with the derivations using the densities. 

The goal is to compute the posterior density of the augmented state given observations, i.e., $\rho(\boldsymbol{u}_{0:N_t} | \boldsymbol{y}_{0:N_t})$. Based on the formulation as a filtering problem, it can be solved sequentially as observations become available. This leads to the filtering distribution, $\rho(\boldsymbol{u}_{n+1} | \boldsymbol{y}_{0:n+1})$. Bayes' theorem then gives us: 
\begin{align} \label{paper_4:eq:posterior}
    \rho(\boldsymbol{u}_{n} | \boldsymbol{y}_{0:n}) =  \frac{\rho(\boldsymbol{y}_{n} | \boldsymbol{u}_{n}) \rho(\boldsymbol{u}_{n} | \boldsymbol{y}_{0:n-1})}{\rho(\boldsymbol{y}_{n}|\boldsymbol{y}_{0:n-1})}.
\end{align}
The problem at hand is to compute equation \eqref{paper_4:eq:posterior} as observations become available. The posterior density is not analytically tractable, so we must resort to numerical approximations. 

We will make use of the particle filter, also referred to as sequential Monte Carlo method, where one aims to sample from the posterior instead of computing it \cite{reich2015probabilistic}. The approximation is performed by creating an ensemble of augmented states (particles) and advancing each augmented state in time using the prior distribution. The posterior is then approximated by the empirical density, made up of $N$ particles,
\begin{align} \label{paper_4:eq:discrete_posterior}
    \rho^{N}(\boldsymbol{u}_{n}| \boldsymbol{y}_{0:n}) = \sum_{i=1}^N w_{n}^{i} \delta\left(\boldsymbol{u}_{n} - \boldsymbol{u}_{n}^{i}\right), 
\end{align}
where $\boldsymbol{u}_{n}^{i}$ represents particle $i$ at time step $n$ and $w_{n}^{i}$ is its corresponding weight. $\delta$ is the Dirac delta function, which gives $w_{n}^{i} \delta(x)= w_{n}^{i}$ for $x=0$ and zero otherwise. $\rho^{N}(\boldsymbol{u}_{n}| \boldsymbol{y}_{0:n})$ can therefore be considered a discrete approximation of the true posterior. The computation of the weights is done by the specific choice of particle filter. A common choice is the bootstrap filter which makes use of importance sampling, originally described in \cite{kitagawa1996monte}. 

The bootstrap filter assimilates data by advancing each particle using the prior distribution and assigning a weight to each. The weights are the normalized likelihoods, computed by evaluating the observation noise density at the residual between the observations and the particles. The weights are then normalized and used to resample the particles using a multinomial distribution with replacement. Note, however, that resampling when a new observation becomes available can lead to poor variablity in the ensemble. Therefore, resampling only occurs when the effective sampling size, $ESS=1/\sum_{i=0}^{N} (w^i_n)^2$, is below a certain threshold, $\lambda_{ESS}$, typically chosen as $N/2$. 



We will refer to the particle filter solution of equation \eqref{paper_4:eq:state_space_model} as the high-fidelity (HF) solution as it is the most accurate solution available. 

The prior distribution is sampled by time stepping in the underlying discretized PDE and adding random noise. For high-dimensional problems, it is generally not feasible to run a particle filter in real-time as several thousands of particles are needed to accurately approximate the posterior, since the true model error is typically unknown. Therefore, reduced order models are often employed to speed up the computations at the cost of accuracy and training time.

\section*{Methodology} \label{paper_4:section:methods}
Here, we present the proposed methodology for real-time data assimilation with particle filters -- the Deep Latent Space Particle Filter (D-LSPF). 

At its heart, we represent the high-fidelity state in a more compact and cheaper to compute latent space and perform the data assimilation in the latent space. We then compute a posterior distribution over the latent state after which we transform back to the high-fidelity space to obtain the high-fidelity posterior. For this, we  employ an autoencoder (AE) to reduce to the latent state, which we combine with a latent time stepping model to advance the latent state. AEs consist of two neural networks: An encoder, $\phi_{\text{enc}}:\boldsymbol{q} \mapsto \boldsymbol{z}$, that reduces the dimension of the data to a latent state, and a decoder, $\phi_{\text{dec}}:(\boldsymbol{z}, \boldsymbol{m}) \mapsto \tilde{\boldsymbol{q}}$, that reconstructs the data. 
The AE is trained by minimizing the loss (MSE) between the input and the reconstruction -- for the parameter dependent cases, we include the parameters in the decoder,  which increases the reconstruction accuracy \cite{makhzani2015adversarial}. The encoder and the decoder are then used to represent equation \eqref{paper_4:eq:state_space_model} in the latent space:
\begin{align} \label{paper_4:eq:latent_state_space_model}
\begin{aligned}[c]
    \boldsymbol{z}_{n} &= f(\boldsymbol{z}_{n-1};\boldsymbol{m}_{n-1}) + \hat{\boldsymbol{\xi}}_{n-1}, \\
    \boldsymbol{m}_{n} &= G(\boldsymbol{m}_{n-1}) + \boldsymbol{\zeta}_{n-1}, \\
    \boldsymbol{y}_n &= h(\phi_{\text{dec}}(\boldsymbol{z}_{n}, \boldsymbol{m}_{n})) + \boldsymbol{\eta}_{n},
\end{aligned}
\quad
\begin{aligned}[c]
     &\hat{\boldsymbol{\xi}}_n\sim P_{\hat{\xi}}(\boldsymbol{\hat{\xi}}_n), \\
     &\boldsymbol{\zeta}_n\sim P_\zeta(\boldsymbol{\zeta}_n), \\
     &\boldsymbol{\eta}_n\sim P_\eta(\boldsymbol{\eta}_n),
\end{aligned}
\end{align}
equation \eqref{paper_4:eq:latent_state_space_model} differs from equation \eqref{paper_4:eq:state_space_model} in three ways:
\begin{itemize}
    \item $\boldsymbol{q}_{n}$ is replaced by $\boldsymbol{z}_{n}$ --  we advance the latent state instead of the high-fidelity state in time;
 
    \item $\boldsymbol{\xi}_n$ is replaced by $\hat{\boldsymbol{\xi}}_n$ -- the latent time stepping model introduces a model error that differs from the high-fidelity model error;
 
    \item $\boldsymbol{q}_{n} = \phi_{\text{dec}}(\boldsymbol{z}_{n})$ is added -- we need to decode the latent state to get synthetic observations in the high-fidelity space.
\end{itemize}

With the augmented latent state, $\boldsymbol{a}_n = (\boldsymbol{z}_{n}, \boldsymbol{m}_{n})$, the high-fidelity posterior density is replaced by the latent posterior density, $\rho(\boldsymbol{a}_{0:N_t} | \boldsymbol{y}_{0:N_t})$. Formulating the problem as a filtering problem, the sequentially defined posterior density is given by: 
\begin{align} \label{paper_4:eq:latent_posterior}
    \rho(\boldsymbol{a}_{n} | \boldsymbol{y}_{0:n}) =  \frac{\rho(\boldsymbol{y}_{n} | \boldsymbol{a}_{n}) \rho(\boldsymbol{a}_{n} | \boldsymbol{y}_{0:n-1})}{\rho(\boldsymbol{y}_{n}|\boldsymbol{y}_{0:n-1})}.
\end{align}
The latent prior density is then computed by:  
\begin{align} \label{paper_4:eq:latent_prior}
    \rho(\boldsymbol{a}_{n}| \boldsymbol{y}_{0:n-1}) = 
    \int \rho(\boldsymbol{a}_{n} | \boldsymbol{a}_{n-1}) \rho(\boldsymbol{a}_{n-1}| \boldsymbol{y}_{0:n-1}) \text{d}\boldsymbol{a}_{n-1},
\end{align}
which is an integral of much lower dimension than the high-fidelity equivalent. The latent likelihood is computed by:
\begin{align} \label{paper_4:eq:latent_likelihood}
    \rho(\boldsymbol{y}_{n} | \boldsymbol{a}_{n}) = \rho_{\eta_n}\left(\boldsymbol{y}_{n} - h(\phi_{\text{dec}}(\boldsymbol{a}_{n}))\right),
\end{align}
which is  faster to evaluate than the high-fidelity equivalent, since $\phi_{\text{dec}}(\boldsymbol{a}_{n}))$ is fast to compute once $f$ and $\phi_{\text{dec}}$ have been trained. 
Eqs. \eqref{paper_4:eq:latent_posterior}, \eqref{paper_4:eq:latent_prior}, and \eqref{paper_4:eq:latent_likelihood} are approximated using the particle filter, as for the high-fidelity equations. The computationally expensive part of the particle filter, namely the time stepping, is performed efficiently in the latent space.

\begin{figure}
 
    \centering
    \includegraphics[width=.65\linewidth]{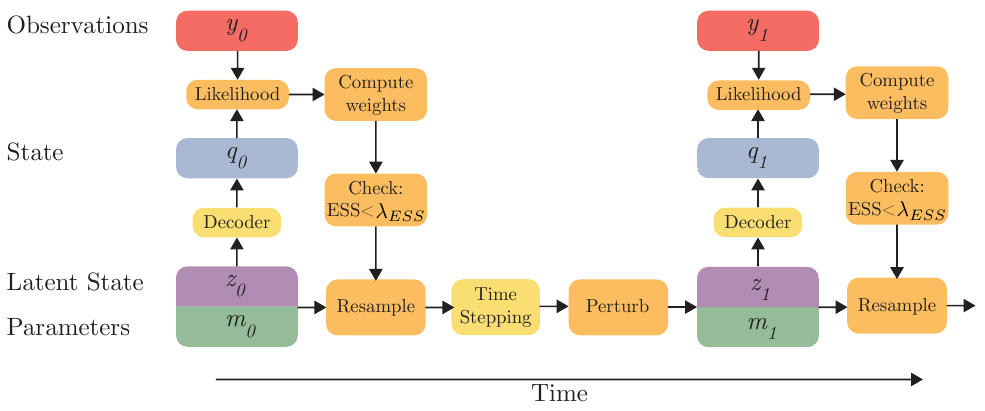}
    \caption{Schematic of the D-LSPF.}
    \label{paper_4:fig:latent_particle_filter}
\end{figure}

Here, we make use of the bootstrap particle filter; other types of particle filter algorithms could possibly be used instead. An outline of the D-LSPF algorithm is shown in Algorithm \ref{paper_4:alg:lsp_filter} and in Figure \ref{paper_4:fig:latent_particle_filter}. 

\begin{algorithm}[ht]
\small
 \KwInput{Trained autoencoder=$(\phi_{\text{enc}}, \phi_{\text{dec}})$, trained time stepping network=$f$, ensemble size=$N$}
    Compute resample threshold, $\lambda_{ESS} = N/2$ \;
    Encode $N$ initial conditions, $\{\boldsymbol{z}^i_0\}_{i=1}^{N} = \{\phi_{\text{enc}}(\boldsymbol{u}^i_0)\}_{i=1}^{N}$ \;
    Initialize weights, $\{w^i_0\}_{i=1}^{N} = \{1/N\}_{i=1}^{N}$  \;

\While{new data, $\boldsymbol{y}_n$, arrives}{
    Time-step latent states, $\{\boldsymbol{z}^i_{n}\}_{i=1}^{N} = \{ f(\boldsymbol{z}^i_{n-1}) + \hat{\xi}_{n-1}^i\}_{i=1}^{N}$ \;
    Decode latent states,  $\{\boldsymbol{u}^i_{n}\}_{i=1}^{N} = \{\phi_{\text{dec}}(\boldsymbol{u}^i_{n})\}_{i=1}^{N}$ \;
    Compute weights, $\{\tilde{w}^i_n\}_{i=1}^{N} = \{\rho_{\eta_{n}}(\boldsymbol{y}_n - h(\boldsymbol{u}^i_{n}))w^i_{n-1}\}_{i=1}^{N}$ \;
    Normalize weights = $\{w^i_n\}_{i=1}^{N} = \left\{\tilde{w}^i_n/\sum_{j=0}^{N}\tilde{w}^j_n\right\}_{i=1}^{N}$ \;
    \If{$1/\sum_{i=0}^{N} (\tilde{w}^i_n)^2 < \lambda_{ESS}$}{
         Resample latent states, $\{\boldsymbol{z}^i_{n}\}_{i=1}^{N}$ with computed weights
    }
}
\KwOutput{Assimilated latent ensemble: $\{\boldsymbol{z}^i_{0:n}\}_{i=1}^{N}$, decoded ensemble: $\{\boldsymbol{u}^i_{0:n}\}_{i=1}^{N} = \{\phi_{\text{dec}}(\boldsymbol{z}^i_{0:n})\}_{i=1}^{N}$}
\caption{D-LSPF (based on the Boostrap particle filter)}
\label{paper_4:alg:lsp_filter}
\end{algorithm}

Note that the speed-up in running the particle filter in the latent space comes at a cost of a training stage and the cost of encoding and decoding (which is fast due to parallel computations on a GPU). This is, however, not a significant drawback as the training takes place offline and the AE and time stepping network can be used numerous times after training.

\subsection*{Latent Space Regularized Autoencoder}
For the D-LSPF to function efficiently, the latent space needs to satisfy certain properties. Firstly, the latent space must be smooth enough: to ensure that the latent space perturbations are meaningful, two states that are close to each other in the high-fidelity space must also be close in the latent space. With smoothness thus defined, we enforce this property using  a prior distribution on the learned latent space in the form of a Wasserstein autoencoder (WAE) \cite{tolstikhin2017wasserstein} using the maximum mean discrepancy (MMD) loss term -- the variational autoencoder (VAE) \cite{kingma2013auto} could serve the same purpose; however, the VAE tends to also smoothen the reconstructions which is undesirable in our settings. 
Secondly, the autoencoder should ensure that latent space trajectories are simple and easy to learn by a time stepping neural network. We achieve this by adding a consistency regularization term, as in \cite{wan2023evolve}, which ensures that the time evolution of the latent state can be modeled by means of an ODE system and thus promotes differentiability, and thereby smoothness, of the time evolution map. 

In summary, for a given training set, $\left\{\boldsymbol{q}_i\right\}_{i=0}^{N_{\text{train}}}$, the complete loss function is given by:   
\begin{align}
\begin{split}
    L_{\text{WAE}}(\phi_{\text{enc}}, \phi_{\text{dec}}) 
    = 
    \frac{1}{N} \sum_{i=1}^{N} \underbrace{\left(\boldsymbol{q}_i - \phi_{\text{dec}}(\phi_{\text{enc}}(\boldsymbol{q}_i)) \right)^2}_{\text{Reconstruction}}
    + \underbrace{\alpha R(\phi_{\text{enc}}, \phi_{\text{dec}})}_{\text{Weight regularization}}
    + \underbrace{\beta \text{MMD}\left(\phi_{\text{enc}}\right)}_{\text{Divergence}}
    + \underbrace{\lambda C(\phi_{\text{enc}}, \phi_{\text{dec}})}_{\text{Consistency}}.
\end{split}
\end{align}

\subsection*{Transformer-Based Dimensionality Reduction}
In our approach, the loss function ensures that the autoencoder, and thereby the latent space, has certain desirable properties. To ensure that the autoencoder can learn a low-dimensional representation and reconstruct it accurately, the architecture also has to be able to handle a multitude of possible high-fidelity states. 

The arguably most common layers for AEs are convolutional and pooling layers \cite{mucke2021reduced, geneva2022transformers, wan2023evolve}. Convolutional layers however tend to have inherent inductive biases and struggle with discontinuous signals, resulting in spurious oscillations. Transformers, originally developed for text processing, have proven effective for image processing in the form of vision transformers (ViT) \cite{dosovitskiy2020image}. These transformers divide images into patches and apply the attention mechanism between each set of patches. Yet, since there is no natural way of reducing or expanding the dimensionality of the data, the ViT has been used to dimensionality reduction tasks only to a limited degree \cite{ovadia2023vito,li2023ib,ran2022transformer, heo2021rethinking}. Importantly, current ViTs have not been integrated with increasing numbers of channels in convolutional layers to represent increasingly complicated features. 

In this section, we extend the vision transformer layer to combine the advantages of convolutional layers (i.e., dimensionality reduction/expansion and increasing/decreasing number of channels) and ViTs (global information, patch processing).

\begin{figure}
    \centering
    \includegraphics[width=0.82\linewidth]{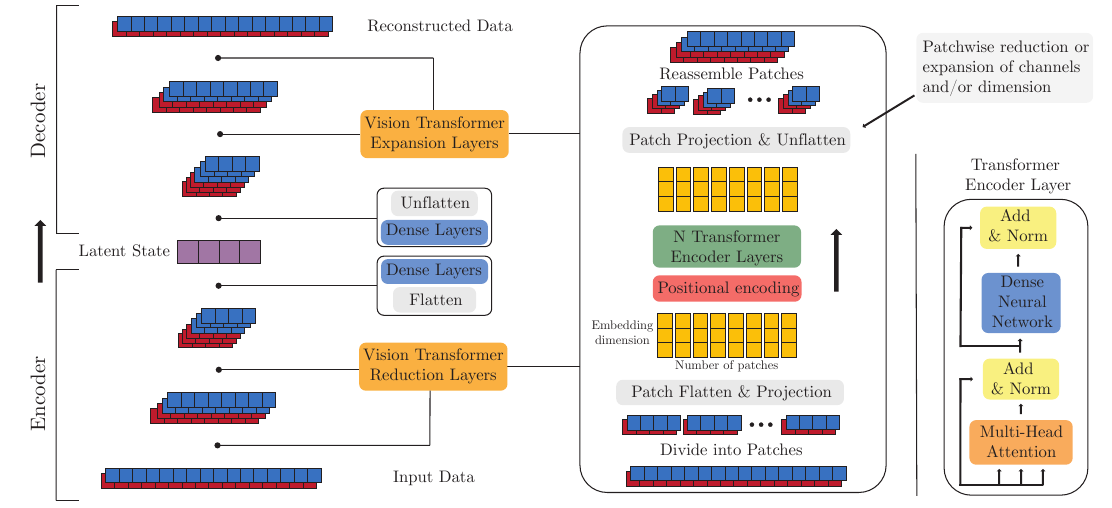}
    \caption{Visualization of the ViT dimensionality reduction/expansion layer.}
    \label{paper_4:fig:transformer_dim_reduction}
\end{figure}

In the proposed layer, the expansion/reduction of channels and dimensions is done on each individual patch. It can be interpreted as a type of domain decomposition, where the reduction/expansion is performed on each subdomain and the communication between subdomains is handled through the attention mechanism. Formally, let the superscript, $l$, denote the $l$'th layer. We divide an input $x^l\in \mathbb{R}^{N_c^l\times N_x^l}$ ($N_c^l$ channels and a spatial dimension of size $N_x^l$), into $p$ patches, $x_1^l, x_2^l, \ldots, x_{p}^l$  of size $N_p^l$.  That is, $x_i^l\in \mathbb{R}^{N_c^l \times N_p^l}$, for all $i$. Then, each patch is flattened and projected onto an $N_e^l$-dimensional embedding space, $e = (e_1, e_2, \ldots, e_p) \in\mathbb{R}^{N_e \times N_p}$. Positional encodings are then added after which the embeddings are passed through a standard transformer encoder layer. Each embedded vector, $e_i$, is projected onto a new dimension of size $N_c^{l+1} N_p^{l+1}$, unflattened, $x_i^{l+1} \in \mathbb{R}^{N_c^{l+1} \times N_p^{l+1}}$ and recombined, $x^{l+1}\in\mathbb{R}^{N_c^{l+1} \times N_x^{l+1}}$. The process is visualized in Figure \ref{paper_4:fig:transformer_dim_reduction}.

\subsection*{Time Stepping}
Once the AE is trained, we can transform high-fidelity trajectories into latent trajectories. To compute the latent trajectories, we next need to perform time stepping in the latent space. For this, we make use of transformers \cite{vaswani2017attention} as they are well suited for modeling physical systems, being able to mimic the structure of multistep time-marching methods \cite{geneva2022transformers}.

Time stepping in the latent space is done by means of a map, $f$, that advances a latent state in time. Adopting the concept of multistep time integrators, we use several previous time steps to predict the next latent state:  
\begin{align}
    \boldsymbol{z}_{n+1} = f(\boldsymbol{z}_{n-k:n} ; \boldsymbol{m}_n),
\end{align}
where $k$ is referred to as the memory. For multiple time steps, we apply the transformer model recursively. Training is done by minimizing the loss function:  
\begin{align*}
    L(f) \hspace{-2pt} = \hspace{-5pt} \sum_{n=k}^{N_t-s}\hspace{-2pt}\sum_{i=1}^s || f^i(\boldsymbol{z}_{n-k:n} ; \boldsymbol{m}_n) - \boldsymbol{z}_{n+1:n+1+i}||^2_2 
   \hspace{-2pt} + \hspace{-2pt}\alpha R(f).
\end{align*}
Here, $R$ is a regularization, $f^i$ means applying $f$ $i$ times, recursively, on the output, and $s$ is the output sequence length. After trajectories are computed in the latent space, high-fidelity trajectories are recovered through the decoder.

\begin{figure}[ht]
    \centering
     
    \includegraphics[width=0.8\linewidth]{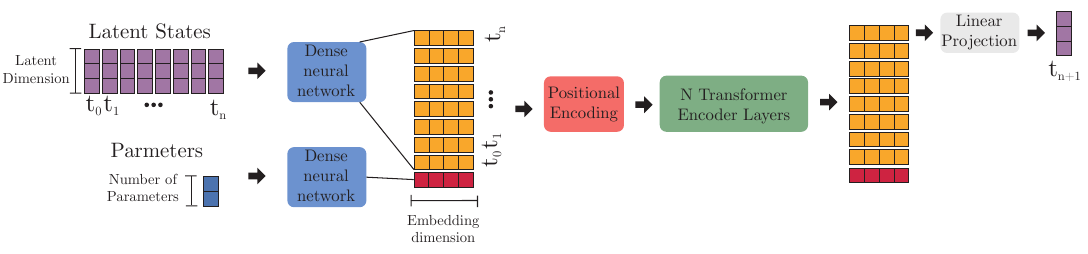}
      
    \caption{Illustration of the transformer model for parameterized time stepping.}
    \label{paper_4:fig:transformer_time_stepping}
      
\end{figure}

In the high-fidelity space,  dynamics are not only dependent on the previous state but also on a set of parameters, and the same applies to the latent dynamics. Including the parameters of interest in the latent space time stepping model can be done in several ways, depending on the specific choice of neural network architecture. We adopt the approach presented in \cite{han2022predicting}, where the parameters are encoded and added to the sequence of states as the first entry:  
\begin{align}
    \left\{g(\boldsymbol{m}_n), \boldsymbol{z}_{n-k}, \boldsymbol{z}_{n-k+1}, \ldots, \boldsymbol{z}_{n}  \right\} \rightarrow
    f\left( \left\{g(\boldsymbol{m}_n), \boldsymbol{z}_{n-k}, \boldsymbol{z}_{n-k+1}, \ldots, \boldsymbol{z}_{n}  \right\}  \right) = \boldsymbol{z}_{n+1},
\end{align}
where $g$ is a parameter encoder that lifts a vector of parameters to the same dimension as the latent state. This efficiently allows attention to be computed between the parameters and the sequence of states. Figure \ref{paper_4:fig:transformer_time_stepping} visualizes the time stepping transformer model. 

\section*{Results and discussion} \label{paper_4:section:results}
We demonstrate the potential and strength of the D-LSPF for a variety of numerical experiments. The first test case serves as a simple benchmark problem. The second test case uses real-world data from an experimental setting, and shows that the D-LSPF can be applied to real-world situations even when trained on simulation data. The last test case is a realistic engineering setting and is used as an ablation study to emphasize the performance of the architectural choices. An overview of the test cases can be found in Table \ref{paper_4:tab:test_case_overview}. For all neural networks, hyperparameter tuning was performed to find the optimal settings. For the alternative methods we compare with, we adopt hyperparameters as chosen in the respective papers when applicable, and performed hyperparameter tuning when not. 

\begin{table}[ht]
\small
    \centering
    \caption{Overview of test cases.}
    \begin{tabular}{lccc} 
    \toprule
     & Burgers & Multi-phase pipeflow & Waves over submerged bar \\ 
    \midrule
    Num Train samples & 1024 & 5000 & 210 \\
    Num Test samples & 20 & 8 & 1 \\
    Parametric & No & Yes & Yes \\
    Simulated observations & Yes & Yes & No \\
    Noise variance & 0.1 & 0.003 & - \\
    Num Sensors & 8 & 9 & 8 \\
    Num Time steps between obs & 30 & 400 & [25, 75] \\
    State DOFs & 256 & 1536 & 1024 \\
    Num States & 1 & 3 & 2 \\
    Num States observed & 1 & 1 & 1 \\
    \bottomrule
\end{tabular}
    \label{paper_4:tab:test_case_overview}
\end{table}

All neural networks were implemented using PyTorch \cite{paszke2019pytorch}. The modified ViT layers are implemented by modifying the code from \url{https://github.com/lucidrains/vit-pytorch}. Training and testing were performed using an Nvidia RTX 3090 GPU and 32 core AMD Ryzen 9 3950X CPU.

All models are trained with the Adam optimizer \cite{kingma2014adam} and a warm-up cosine annealing learning rate scheduler. Gradient clipping was applied when training the transformers. The states and parameters were transformed to be between 0 and 1 before being passed to the autoencoder. The time stepping networks are trained without teacher forcing, and with a limited unrolling. The number of time steps to unroll was treated as a hyperparemeter.  

\subsection*{Viscous Burgers Equation}
The first test case is the viscous Burgers equation:
\begin{align} \label{paper_4:burgers}
\begin{split}
    &\partial_t q(x, t) = \nu \partial_{xx}q(x, t) - q(x, t) \partial_x q(x, t), \\
    &q(0, t) = q(L, 0) = 0, \\
    &q(x, 0) = Q \sin\left( \frac{2\pi x}{L} \right),
\end{split}
\end{align}
with $x\in [0,L]$, $L=2$, $\nu=1/150$, and $Q \sim \text{U}[0.5, 1.5]$. We only perform state estimation so neither the AE nor the time stepping NN receives any parameters as input. The observations used for the data assimilation are simulated, as well as the training data. We add normally distributed noise with a standard deviation of 0.1. equation \eqref{paper_4:burgers} is discretized using a second-order finite difference scheme in space and a Runge-Kutta 45 method in time. We consider $t\in [0, 0.3]$ with a step size of 0.001, resulting in 300 time steps. For the data assimilation, we test on a case where the state is observed at 8 spatial locations, $x=(0.0 , 0.286, 0.571, 0.857, 1.143,1.429, 1.714, 2.0)$, $N_y=8$, at every 10th time step. The latent dimension in the D-LSPF is chosen to be 16. 

We compare the D-LSPF with 100 and 1000 particles with the Reduced-Order Autodifferentiable Ensemble Kalman Filter (ROAD-EnKF) \cite{chen2023reduced} with a latent dimension of 40. The ROAD-EnKF is trained on the same training data as the D-LSPF with full access to the entire states in space and time. All methods are evaluated on 20 different simulated solutions, with $Q \sim \text{U}[0.5, 1.5]$. We compare the performance by computing the Root-Mean-Square Error (RMSE) and the averaged RMSE of the 2nd, 3rd, and 4th moment of the state ensemble, referred to as the Average Moment RMSE (AMRMSE). The AMRMSE measures how accurately the distributional information of the posterior is approximated and therefore how accurately uncertainty is quantified. Moreover, we also present the negative log-likelihood (NLL) with respect to the high-fidelity posterior.

In Table \ref{paper_4:tab:burgers_results}, the results for the test case are shown. The D-LSPF shows superior performance with respect to AMRMSE and NLL by one order of magnitude, suggesting that the D-LSPF quantifies the uncertainty in a more accurate way compared to ROAD-EnKF for this case. For the mean state estimation, the D-LSPF also performs 3.75 times better. Regarding timing, the D-LSPF with 100 particles and the ROAD-EnKF are comparable using GPUs, while the ROAD-EnKF is slightly faster using a CPU. 

\begin{figure*}
     \centering
      \begin{subfigure}[b]{0.32\linewidth}
         \centering
         \includegraphics[width=1.\textwidth]  {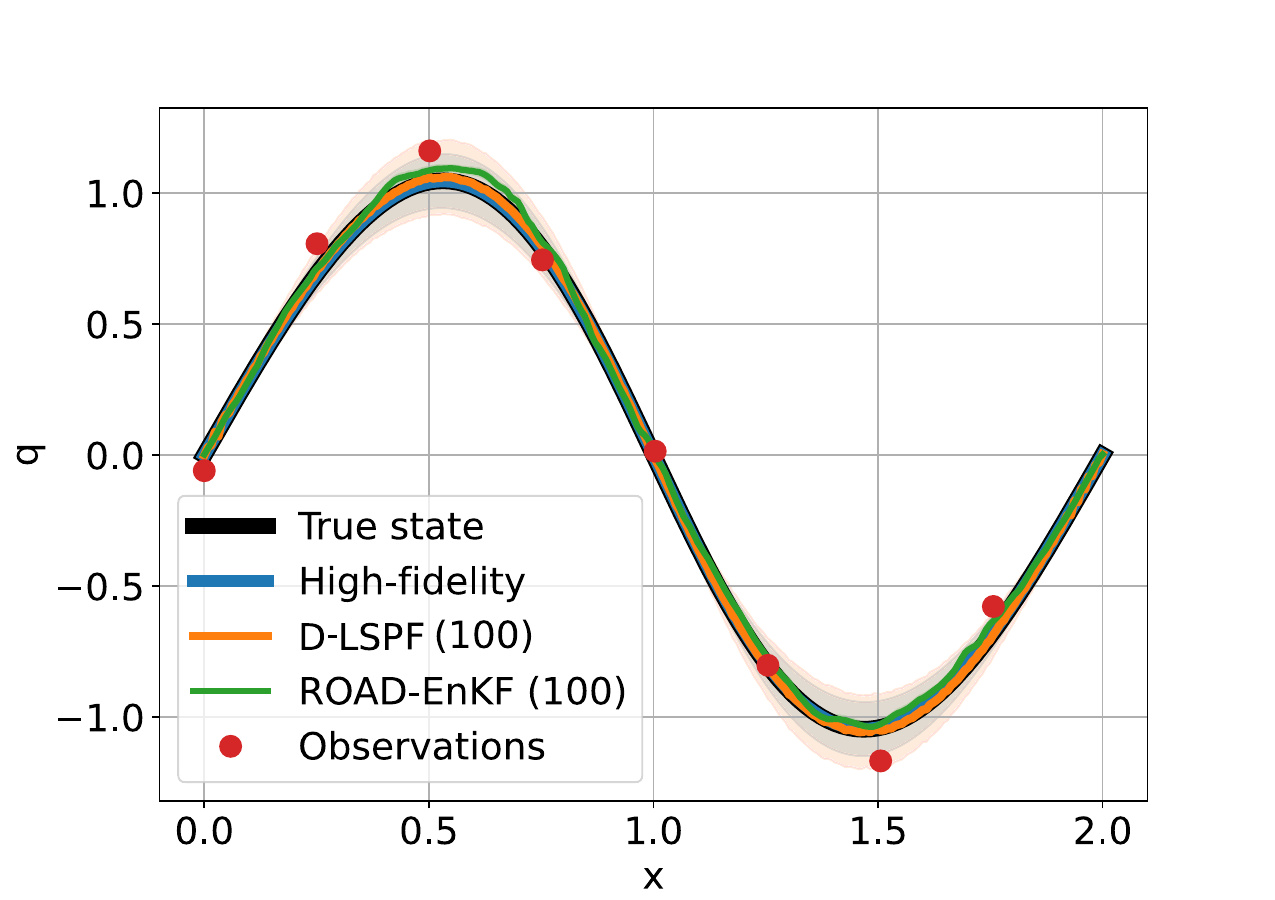}
         \caption{$t=0.03$}
         \label{paper_4:fig:burgers_comparison_30}
     \end{subfigure}
     \begin{subfigure}[b]{.32\linewidth}
         \centering
         \includegraphics[width=1.\textwidth]  {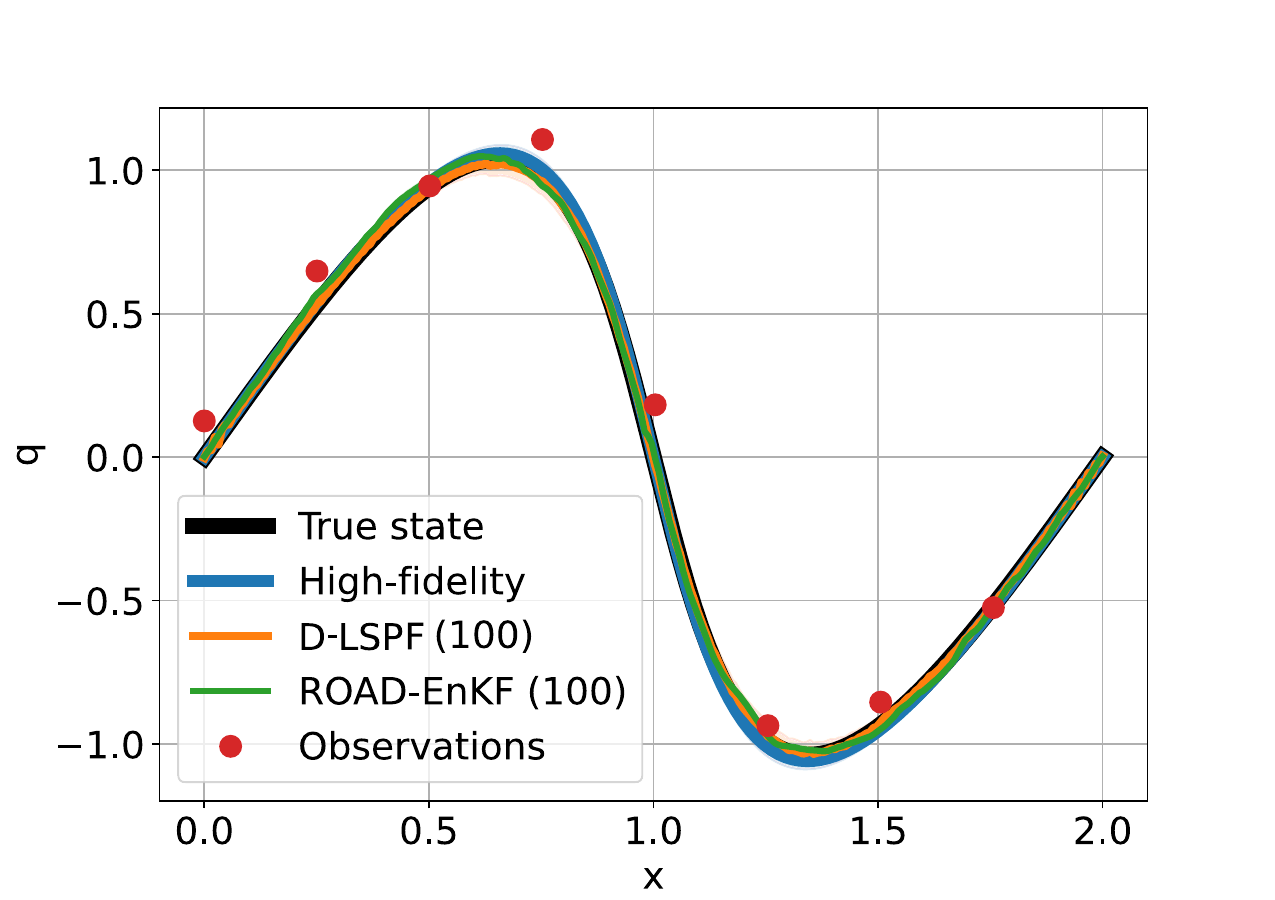}
         \caption{$t=0.15$}
         \label{paper_4:fig:burgers_comparison_150}
     \end{subfigure}
     \begin{subfigure}[b]{.32\linewidth}
         \centering
         \includegraphics[width=1.\textwidth]  {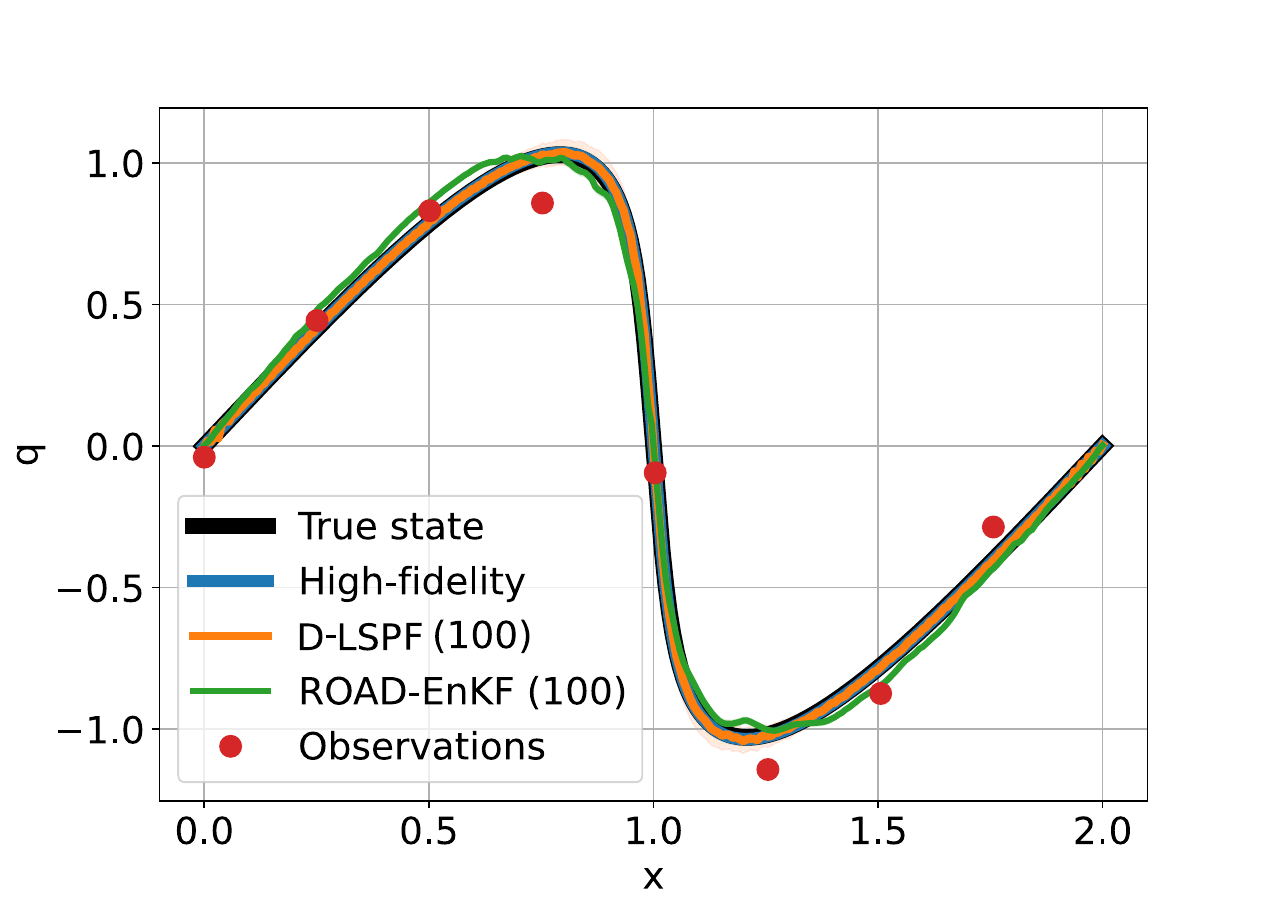}
         \caption{$t=0.29$}
         \label{paper_4:fig:burgers_comparison_289}
     \end{subfigure}
    \caption{State estimation for the viscous Burgers equations with observations every 10 time step using the high-fidelity particle filter, the D-LSPF and the ROAD-EnKF. The high-fidelity particle filter was run with 1000 particles and the D-LSPF and ROAD-EnKF were run with 100 particles. we see the state estimation for the using the D-LSPF, ROAD-EnKF, and the high-fidelity model for case 2. It is clear that all methods approximates the state well. However, the ROAD-EnKF is visibly worse at the end.  }
    \label{paper_4:fig:burgers_state_results}
\end{figure*}

Lastly, in Figure \ref{paper_4:fig:burgers_state_results}, we show state estimation results at three different time points. It is clear that the uncertainty bands shrink as time passes and more observations become available as expected. The ROAD-EnKF state estimation is visually slightly worse than the D-LSPF and the high-fidelity particle filter. 
\begin{table}
\small
    \centering
    \caption{The viscous Burgers equations. In parenthesis is the number of particles. For the high-fidelity (HF) particle filter, 30 CPUs were used in parallel. For the rest of the methods a single CPU or GPU was used. The AMRMSE is computed by comparing the surrogate model ensembles with the HF particle filter solution. }
    \begin{tabular}{l|ccccc} 
    \toprule
    &   RMSE $\downarrow$ &   AMRMSE $\downarrow$ & NLL $\downarrow$ & GPU $\downarrow$ &   CPU $\downarrow$ \\
    \midrule
       HF(1000)&    $9.0\cdot 10^{-3}$ &  - & & - &    22.7s \\
    \midrule
       D-LSPF(100)&   $\mathbf{1.6\cdot 10^{-2}}$ &  $1.3\cdot 10^{-4}$  & \textbf{0.38} &  \textbf{0.6s} &     1.5s \\
       D-LSPF(1000) &   $\mathbf{1.6 \cdot 10^{-2}}$ &   $\mathbf{1.1 \cdot 10^{-4}}$ & 0.42 &  1.2s &    13.4s  \\
       ROAD-EnKF&    $6.0\cdot 10^{-2}$ &   $1.2\cdot 10^{-3}$ & 4.73 &  \textbf{0.6s} &   \textbf{0.8s} \\
    \bottomrule
\end{tabular}
\label{paper_4:tab:burgers_results}
\end{table}

\subsection*{Harmonic wave generation over a submerged bar} \label{paper_4:section:wave_test_case}
\begin{figure}
    \centering
    \includegraphics[width=.4\textwidth]  {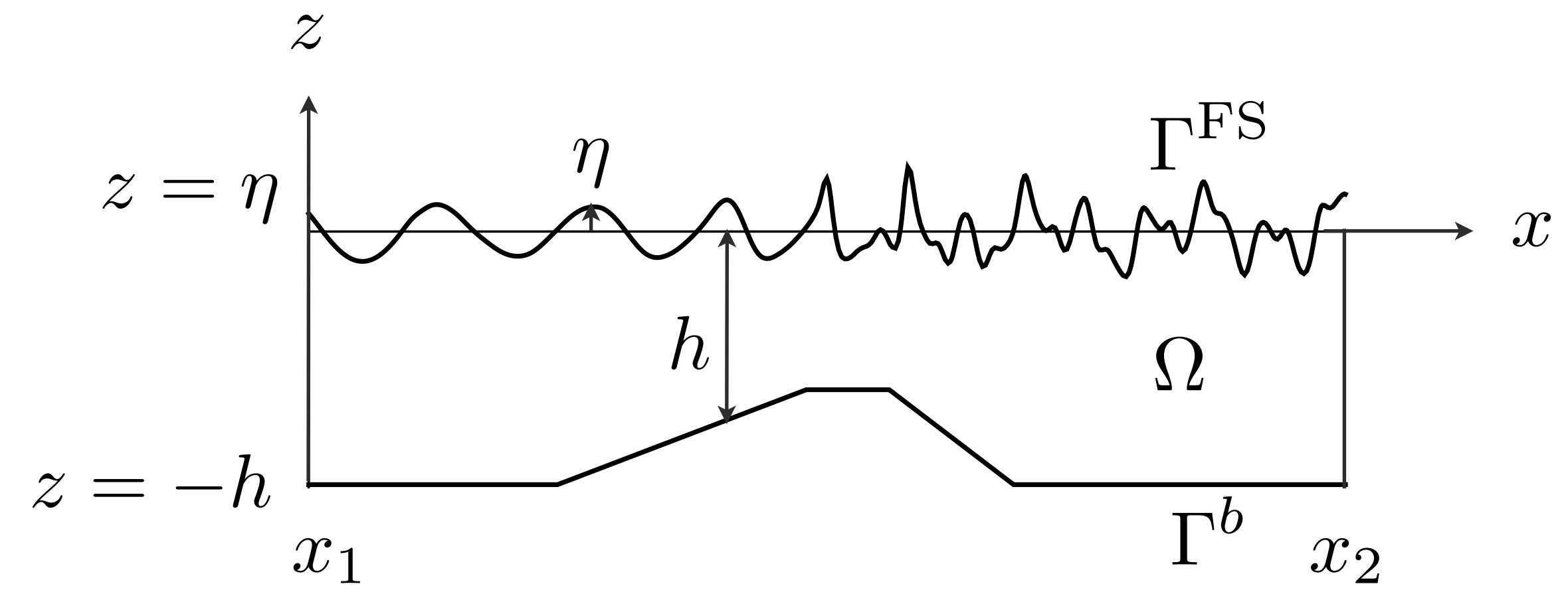}
    \caption{Wave tank setup and physical variables. Figure comes from \cite{engsig2016stabilised}.}
    \label{paper_4:fig:wave_setup}
\end{figure}

In this test case, the data comes from a real-world experiment \cite{beji1994numerical}. The setting is a 25m long and 0.4m tall wave tank, with waves being generated from the left side, traveling to the right. At the seabed of the tank, a 0.3m tall submerged bar is placed, see Figure \ref{paper_4:fig:wave_setup}. Eight sensors measure the surface height of the water at $x=(4, 10.5, 13.5, 14.5, 15.7, 17.3, 19.0, 21.0)$. For the state and parameter estimation, we aim to reconstruct the surface elevation, the velocity potential, and the height of the submerged bar. A similar study was conducted in \cite{bigoni2016efficient}, where the uncertainty of the water wave height was quantified given random perturbations on the seabed; in \cite{bigoni2016efficient} however only uncertainty of the forward problem was considered, whereas we solve the inverse problem with uncertainty quantification, given the observations. For the neural network surrogate model, we only consider the surface variables $\eta$ and $\Tilde{\phi}$, as the state and the height of the submerged bar is the parameter of interest. This highlights an important advantage of using a non-intrusive surrogate model, as it becomes possible to only model the relevant quantities of interest, bypassing the computations of the velocity potential in the vertical direction. 

To generate the training data, we model the setup using the fully nonlinear water wave model for deep fluids as described in \cite{engsig2016stabilised}. The problem is modeled in 2D by means of a set of 1D PDEs for the free surface boundary conditions, together with a 2D Laplace problem in the full domain. Let $x$ be the horizontal component and $z$ the vertical component, see Figure \ref{paper_4:fig:wave_setup}. The velocity potential, $\phi:(x,z,t)\mapsto \phi(x,z,t)$, is the scalar function defined on the whole 2D domain, and the free surface elevation, $\eta:(x,t)\mapsto \eta(x,t)$, is defined only on the 1D surface. The free surface boundary conditions can be expressed in the so-called Zakharov form \cite{zakharov1968stability}, modeled by two 1D PDEs -- the wave height, $\eta$, and the velocity potential, $\Tilde{\phi}$:
\begin{align}\label{paper_4:eq:FSproblem}
\begin{split}
\frac{\partial \eta}{\partial t}  &= -\boldsymbol{\nabla}\eta\cdot\boldsymbol{\nabla}\tilde{\phi}+\tilde{w}(1+\boldsymbol{\nabla}\eta\cdot\boldsymbol{\nabla}\eta), \\
\frac{\partial \tilde{\phi}}{\partial t} &= -g\eta - \frac{1}{2}\left(\boldsymbol{\nabla}\tilde{\phi}\cdot\boldsymbol{\nabla}\tilde{\phi}-\tilde{w}^2(1+\boldsymbol{\nabla}\eta\cdot\boldsymbol{\nabla}\eta)\right).
\end{split}
\end{align}
equation \eqref{paper_4:eq:FSproblem} is defined on the surface part of the domain, $\Gamma^{\textrm{FS}}$. $\tilde{w}=\partial_z\phi|_{z=\eta}$ and $\tilde{\phi}=\phi|_{z=\eta}$ are the surface parts of the functions that are defined on the 2D domain and $\Gamma^{\textrm{FS}}$ is the free surface, as shown in Figure \ref{paper_4:fig:wave_setup}. The velocity potential on the domain is modeled by the 2D Laplace problem, via the $\sigma$-transform:
\begin{align} \label{paper_4:eq:laplace}
\begin{split} 
\nabla^{\sigma}( K(x;t)\nabla^{\sigma}\phi) &= 0,\quad \textrm{in} \quad \Omega^c, \\
    \phi &=\tilde{\phi}, \quad z=\eta \quad\textrm{on}\quad \Gamma^{FS} \\
    {\bf n}\cdot \nabla\phi &= 0, \quad  z=-h({x,y})\quad \textrm{on}\quad \Gamma^b.
\end{split}
\end{align}
where $\sigma = (z+h(x))d(x, t)^{-1}$, $\Omega^c=\{(x,\sigma)|0\leq \sigma \leq 1\}$, and
\begin{align}
    K(x, t) = \begin{bmatrix}
        d & -\sigma \partial_{x}\eta \\ 
        -\sigma \partial_{x}\eta & \frac{1 + (\sigma \partial_{x}\eta)^2}{d}
    \end{bmatrix}.
\end{align}
We use the spectral element method, as described in \cite{engsig2016stabilised}, for the discretization of the equations. We use 103 elements in the horizontal direction and 1 element in the vertical direction, both with 6th order polynomials, to generate the training data. The equations are solved with a step size of 0.03535 with $t\in[0, 42.42]$, resulting in 1200 time steps, and the bar height is uniformly sampled between 0.1 and 0.325. The states are interpolated onto a regular grid of 512 points.

Sensor observations from the experiment are available at a time frequency of 0.03535s, which was also chosen as the step size for the simulations. To demonstrate how the D-LSPF performs with varying time intervals between the observations, we show the results for sensor observations at every 25th and 75th time step, corresponding to every 0.884s and 2.651s, respectively. We refer to these two settings as case 1 and 2. We only observe the wave height and not the velocity potential. Since we deal with real-world data, the true full state is not available. Therefore, we measure the accuracy against the full time series of observations, showcasing that the D-LSPF can accurately estimate the state between observations. We do, however, also compare the results with a high-fidelity simulation with the true bar height. Furthermore, we present the accuracy of the bar height estimates. 

\begin{table*}[ht]
    \small
\centering
\caption{Results for the harmonic wave generation test case using the D-LSPF and the ROAD-EnKF with 100 and 1000 particles. Timings are measured using a single GPU. The PICP is computed using the 2.5th and the 97.5th percentile. An upward pointing arrow means larger values are better and a downward pointing arrow means lower values are better. "S-" and "P-" refer to the state and parameters, respectively.}
    

        

\begin{tabular}{l|cccc|cccc} 
    
    \toprule
    &\multicolumn{4}{c}{Case 1 -- Every 25 time step} &\multicolumn{4}{c}{Case 2 -- Every 75 time step} \\
    \midrule
    & S-RRMSE $\downarrow$ & S-PICP $\uparrow$ &P-RRMSE $\downarrow$ &  Time $\downarrow$ & S-RRMSE $\downarrow$ & S-PICP $\uparrow$ &P-RRMSE $\downarrow$ &  Time $\downarrow$ \\

    \midrule
    D-LSPF(100) &   $\mathbf{3.6 \cdot 10^{-1}}$  &    $4.4 \cdot 10^{-1}$ &   $\mathbf{4.4 \cdot 10^{-3}}$ &   \textbf{1.5s} & 
    $4.3 \cdot 10^{-1}$ &   $4.9 \cdot 10^{-1}$ &   $5.3 \cdot 10^{-3}$ &  \textbf{1.4s} \\
    D-LSPF(1000) &   $3.8 \cdot 10^{-1}$ &   $3.9 \cdot 10^{-1}$ &   $4.6 \cdot 10^{-3}$ &   10.5s &  
    $\mathbf{4.0 \cdot 10^{-1}}$ &   $5.6 \cdot 10^{-1}$  &   $\mathbf{4.9 \cdot 10^{-3}}$ &   10.0s \\
    ROAD-EnKF(100) &   1.1 &   $5.0 \cdot 10^{-1}$ & - &   5.4s &   
    1.0 &  $\mathbf{5.8 \cdot 10^{-1}}$& - &    5.1s\\
     ROAD-EnKF(1000) &   1.0 &   $\mathbf{5.5 \cdot 10^{-1}}$ & - &   31.3s&   
     1.0 &   $\mathbf{5.8 \cdot 10^{-1}}$ & - &   29.9s  \\
    \bottomrule
\end{tabular}
\label{paper_4:tab:wave_results}
\end{table*}
    
We compare the D-LSPF with the ROAD-EnKF \cite{chen2023reduced}, where we train the ROAD-EnKF model on the same simulated data as the D-LSPF with full access to the states in space and time. To deal with the multiple states, wave height and velocity potential, we introduce a slight modification in the decoder network in the ROAD-EnKF compared to \cite{chen2023reduced}, by ensuring that the Fourier decoder networks outputs data with two channels. The latent dimension is chosen to be 8 for the D-LSPF and 40 for the ROAD-EnKF. The neural network architectures for the modified ROAD-EnKF model have been chosen through hyperparameter tuning. 

\begin{figure*}[ht]
    \centering
    \begin{subfigure}{.45\linewidth}
        \centering
        \includegraphics[width=.8\textwidth]  {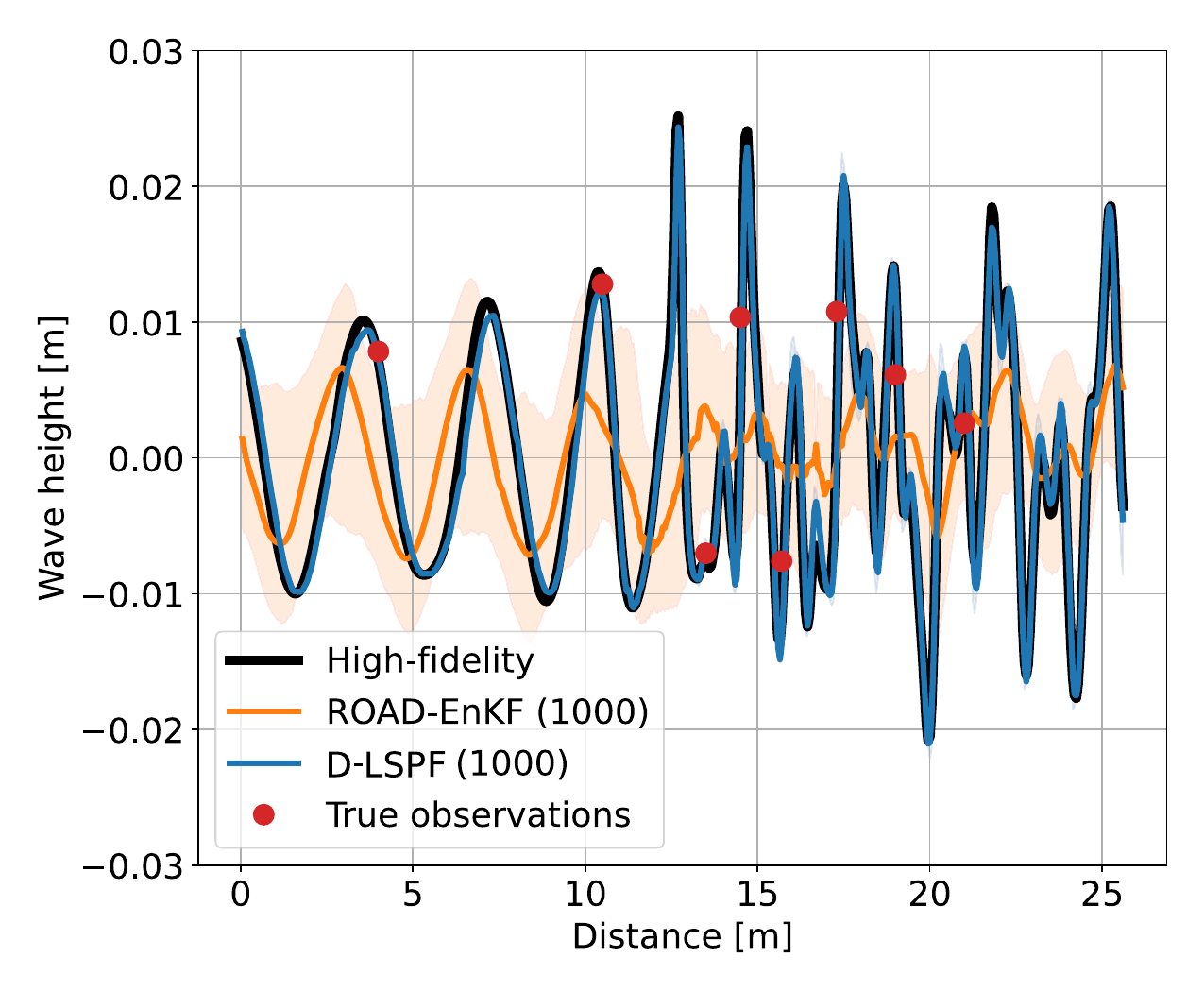}
        \caption{Wave height estimation, $\eta \pm 2\sigma$.}
    \end{subfigure}
    \begin{subfigure}{.45\linewidth}
        \centering
        \includegraphics[width=.8\textwidth]  {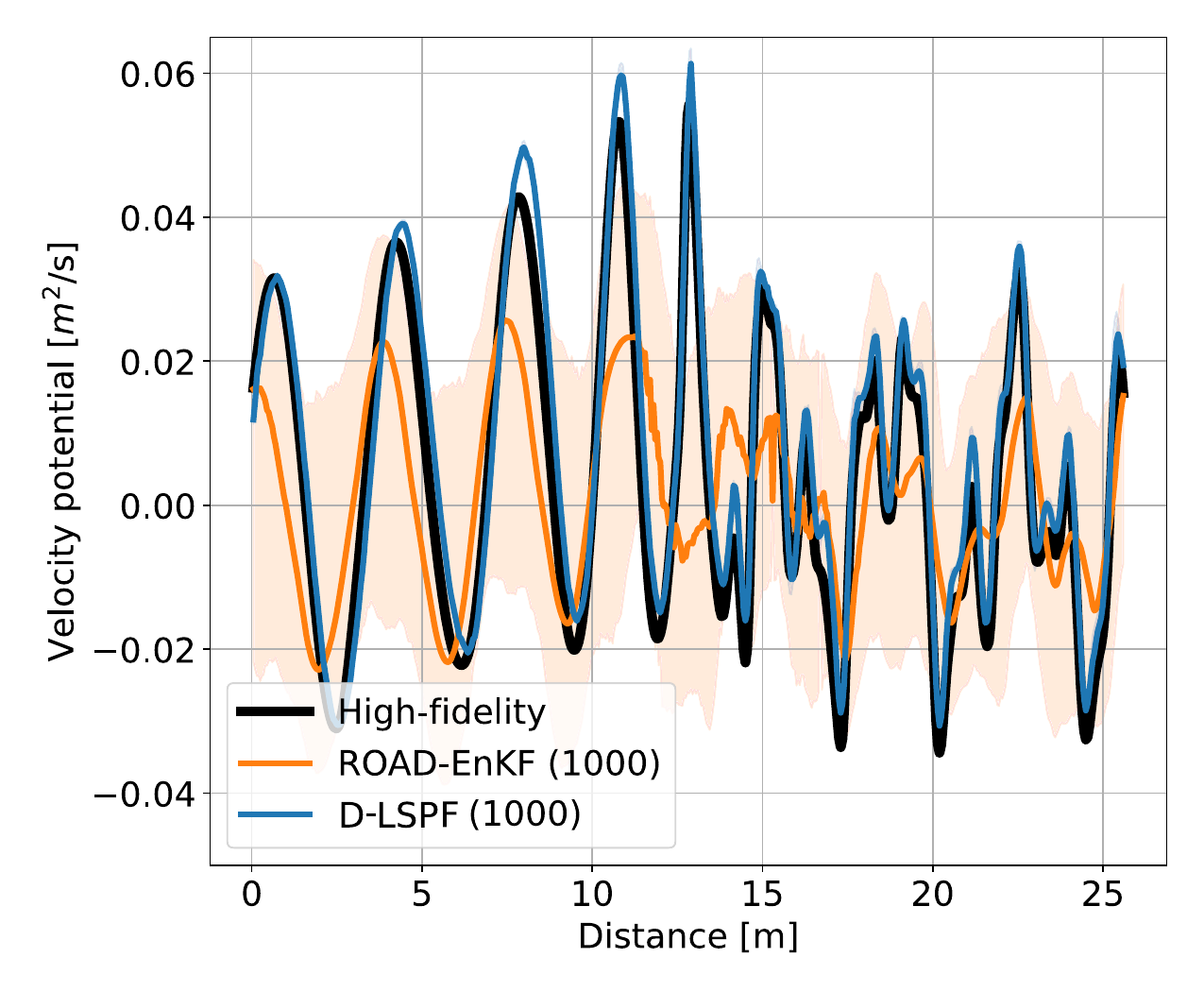}
        \caption{Velocity potential estimation, $\tilde{\phi} \pm 2\sigma$.}
    \end{subfigure}
    \caption{State estimation at t=40s for the harmonic wave test case with observations every 75 time steps. Note that only wave height is observed and not velocity potential.}
    \label{paper_4:fig:wave_state_estimation}
\end{figure*}

\begin{figure*}
    \centering
    \begin{subfigure}{.48\linewidth}
        \centering
        \includegraphics[width=.85\textwidth]  {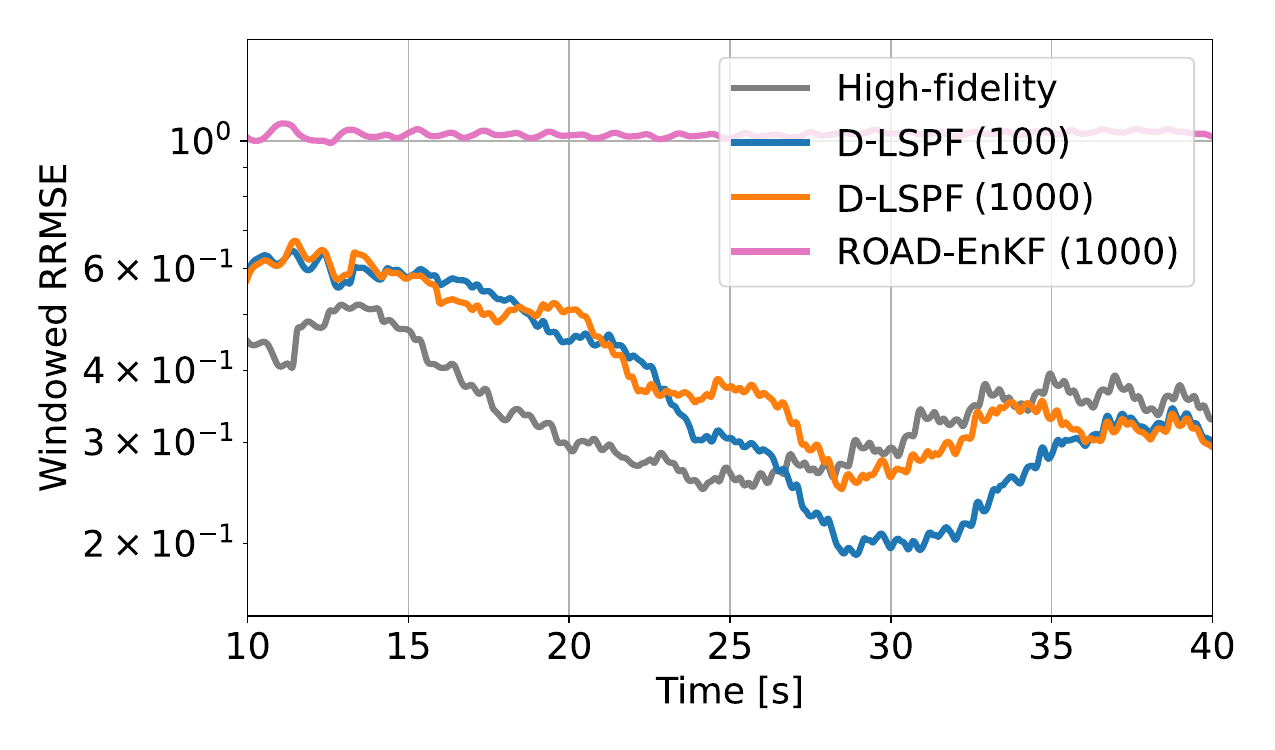}
        \caption{Observations every 25 time step.}
    \end{subfigure}
    \begin{subfigure}{.48\linewidth}
        \centering
        \includegraphics[width=.85\textwidth]  {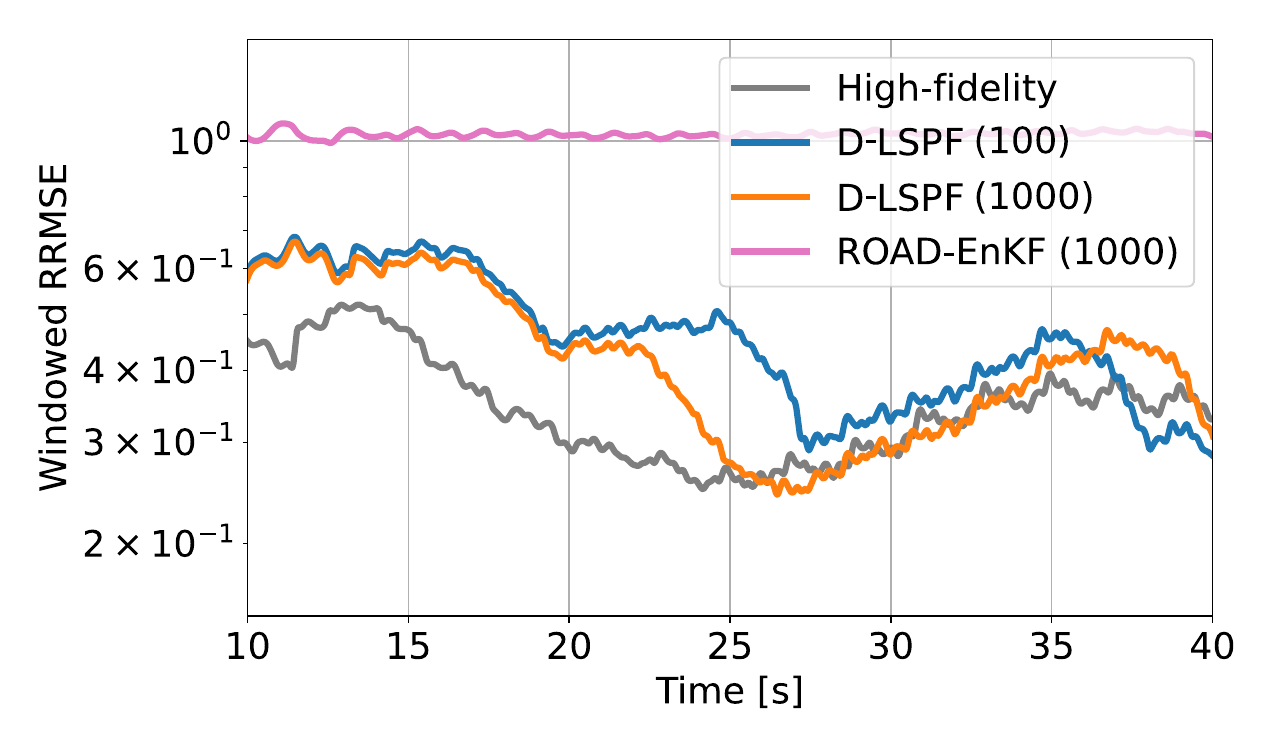}
        \caption{Observations every 75 time step.}
    \end{subfigure}
    \caption{Windowed RRMSE for the harmonic wave test case with a window size of 75 time steps.}
    \label{paper_4:fig:windowed_RRMSE}
\end{figure*}

Table \ref{paper_4:tab:wave_results} contains the Relative RMSE (RRMSE), probability interval coverage percentage (PICP), and timings for the D-LSPF and the ROAD-EnKF in both variations of the test case using 100 and 1000 particles. The D-LSPF clearly performs best with respect to the state estimation with an improvement of one order of magnitude, for both 100 and 1000 particles. For the PICP, the ROAD-EnKF does slightly better, however, inspecting Figures \ref{paper_4:fig:wave_state_estimation}a and \ref{paper_4:fig:wave_state_estimation}b, we see that the ROAD-EnKF has large uncertainty intervals while being quite inaccurate on average compared to the D-LSPF. In general, the PICP is less relevant when the RRMSE is bad. This is further highlighted in Figure \ref{paper_4:fig:windowed_RRMSE}a and \ref{paper_4:fig:windowed_RRMSE}b, plotting the windowed RRMSE versus time. The windowed RRMSE measures the RRMSE in a time window in order to show how the state estimation improves when more observations become available. The D-LSPF converges and even surpasses the high-fidelity simulation, showcasing how assimilating observations impactfully improves accuracy. We also notice for both the D-LSPF and ROAD-EnKF that there are only minor differences between using 100 and 1000 particles. Furthermore, the RRMSE only varies slightly between the two cases, suggesting that both methods are stable with respect to observation frequency.  

Besides the accuracy, Table \ref{paper_4:tab:wave_results} also notes the computation time. In both cases, both the D-LSPF and the ROAD-EnKF are faster than real-time, as the data assimilation takes place over 40s in physical time and the D-LSPF and ROAD-EnKF computation times vary between 1.4s and 29.9s. In general, the D-LSPF is between 3 and 4 times faster than the ROAD-EnKF. For comparison, a single high-fidelity model simulation takes on average some 1062s.  Hence, running the particle filter using the high-fidelity model on 30 CPU cores, assuming no overhead associated with the parallelization, would take approximately 3542s with 100 particles and 35420s for 1000 particles, yielding a speed-up of 2345 for 100 particles and 3376 for 1000 particles on a GPU when using the D-LSPF. When deploying the D-LSPF and running it as the data comes in, it is not possible to perform the data assimilation task faster than the arrival of observations. However, the timings show that the method  assimilates the data without any delay. 

\begin{figure*}[ht]
    \centering
    \begin{subfigure}{.45\linewidth}
        \centering
        \includegraphics[width=1.\textwidth]  {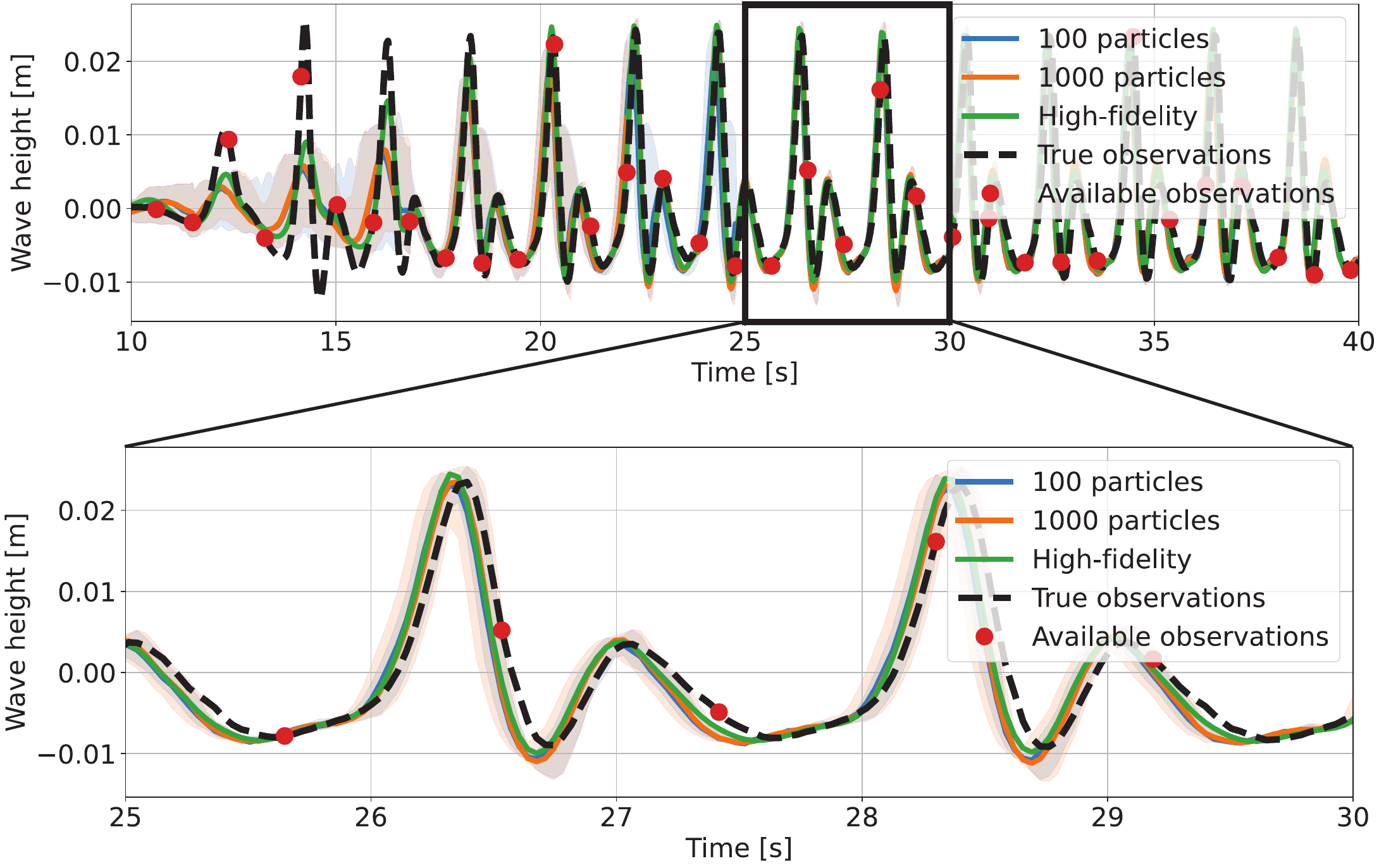}
        \caption{Sensor located at $13.5m$.}
    \end{subfigure}
    \begin{subfigure}{.45\linewidth}
        \centering
        \includegraphics[width=1.\textwidth]  {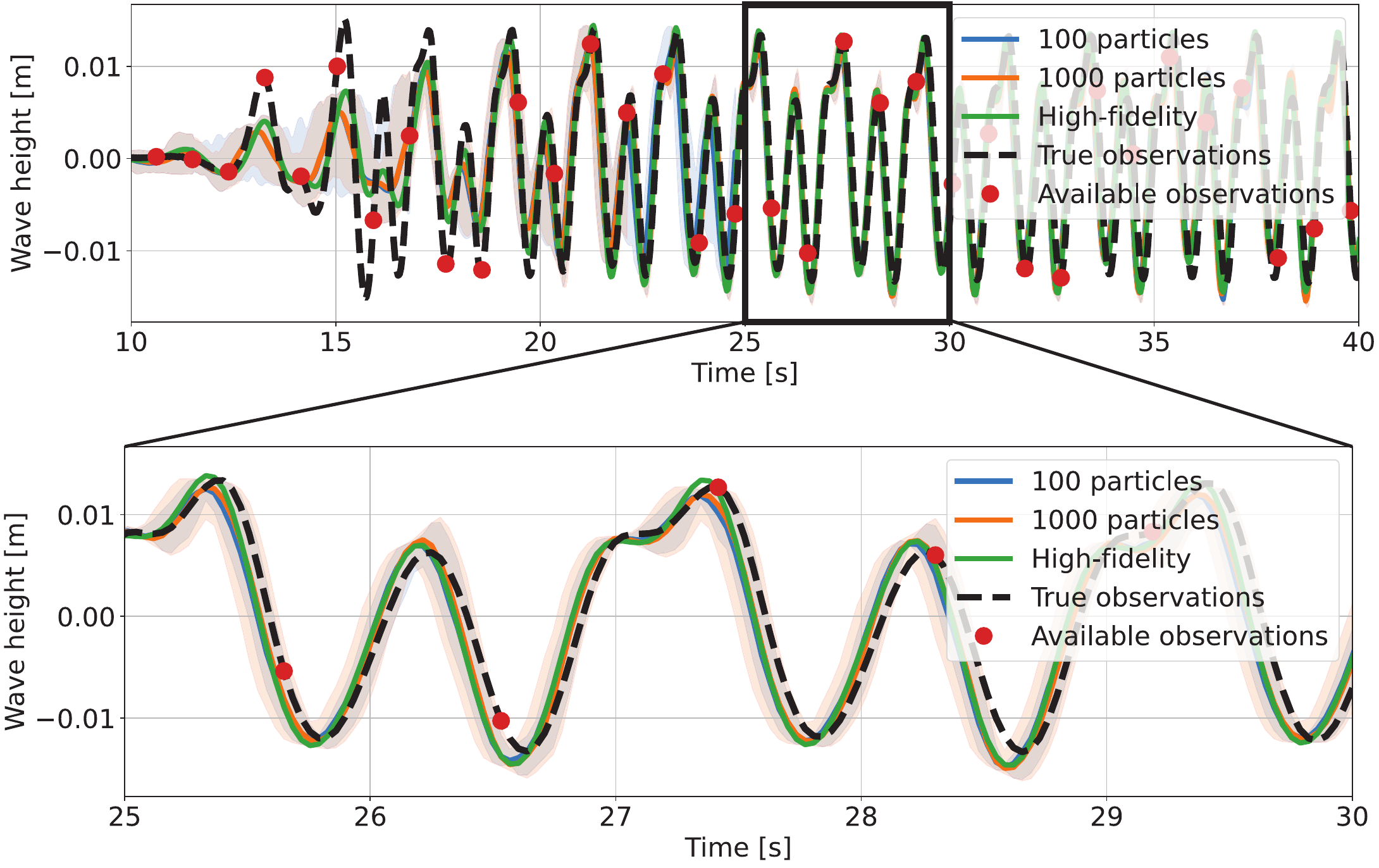}
        \caption{Sensor located at $14.5m$.}
    \end{subfigure}
    \caption{State estimation at two sensor locations computed with the D-LSPF with 100 and 100 particles compared with a high-fidelity simulation with the true bar height and the true sensor data.}
    \label{paper_4:fig:wave_sensor_3_4}
\end{figure*}

Figure \ref{paper_4:fig:wave_sensor_3_4} shows the quality of the D-LSPF estimates of the state and the sensor locations between observations. Furthermore, the difference between using 100 and 1000 particles is negligible for the state estimates. However, when zooming in, it becomes clear that using more particles result in better uncertainty intervals. 
 
\begin{figure*}
    \centering
    \begin{subfigure}{.4\linewidth}
        \centering
        \includegraphics[width=1.\textwidth]  {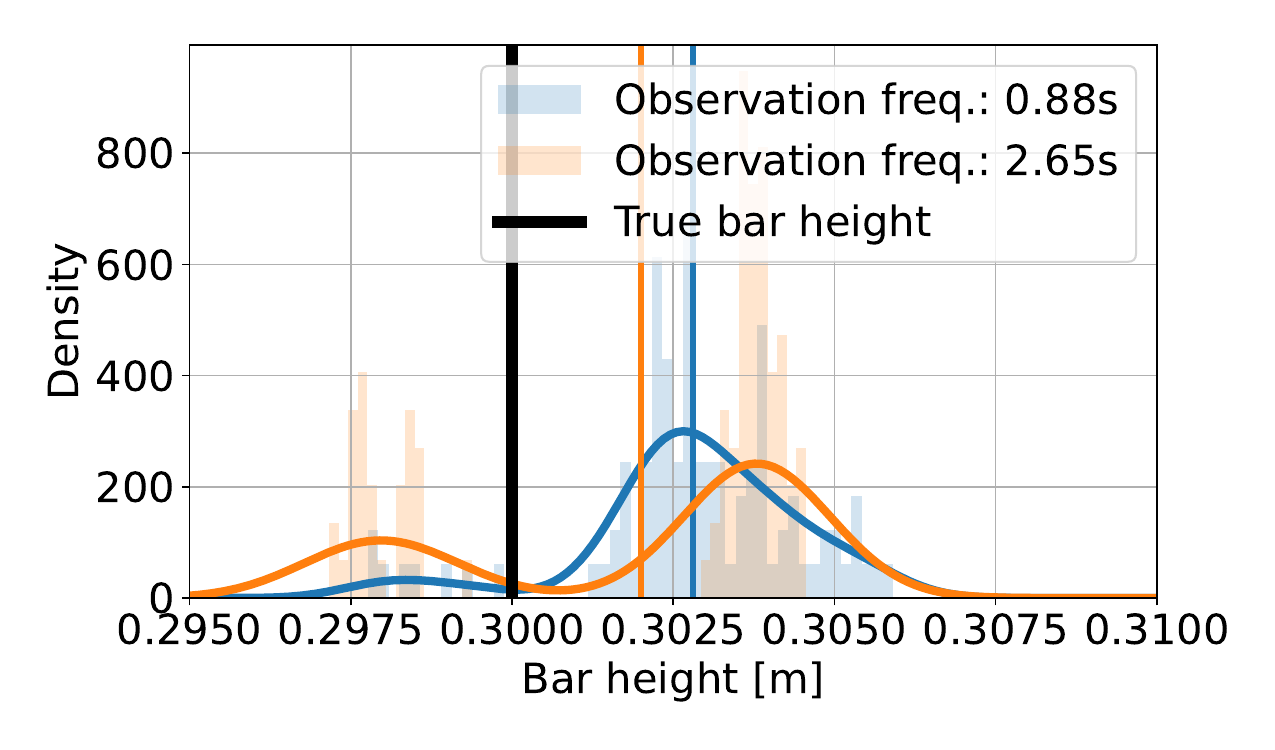}
        \caption{D-LSPF with 100 particles.}
    \end{subfigure}
    \begin{subfigure}{.4\linewidth}
        \centering
        \includegraphics[width=1.\textwidth]  {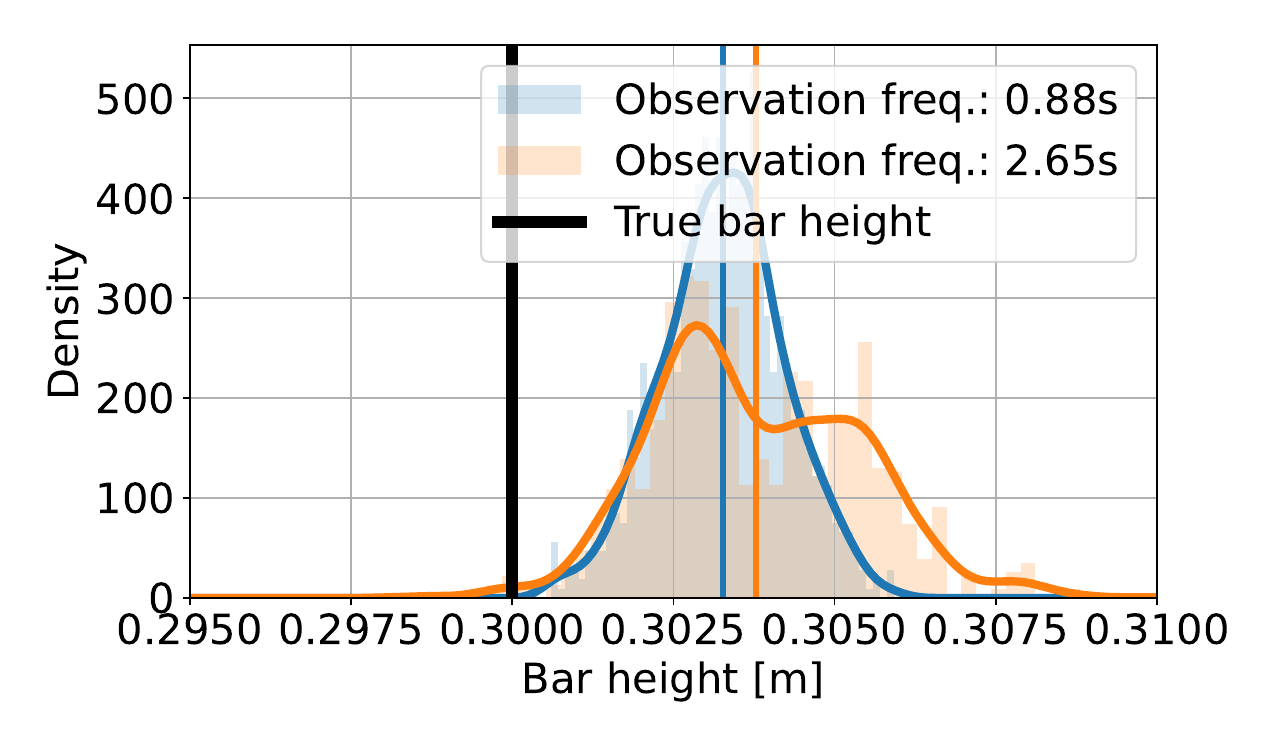}
        \caption{D-LSPF with 1000 particles.}
    \end{subfigure}
    \caption{Posterior distribution over the bar height at $t=40s$ for varying observation frequency and number of particles. The density outlines are computed using kernel density estimation based on the posterior particles. Vertical lines represent the mean of the distributions.}
    \label{paper_4:fig:bar_height_hist}
\end{figure*}

Lastly, in Figure \ref{paper_4:fig:bar_height_hist}, the posterior distributions of the bar height for varying numbers of particles and observation frequency are shown. While the average bar height estimates are very similar for the 100 and 1000 particles, the distributions change from being multimodal to unimodal.


\subsection*{Multi-phase leak localization}
Here, we consider leak localization in a two-phase pipe flow for liquid (water) and air, and we assume that they are mixed. We consider a case where only a few sensors are placed along a 5km long pipe, measuring pressure. The system is in a steady state when a leak occurs, where we then use the particle filter to compute a distribution over the leak location, leak size, liquid holdup, mixture velocity, and pressure. This problem is similar to the setup in \cite{uilhoorn2014particle}, where single-phase CO$_2$ was considered, but state estimation was only performed in a non-leakage case. 

The state for data assimilation and our surrogate model is the primitive state, consisting of the liquid holdup, pressure, and mixture velocity, $q=(\alpha_l, p, u_m)$. Both training, testing, and observations data are simulated and noise is artificially added to the observations with a standard deviation of $0.01$. We use the homogeneous equilibrium model (HEM) \cite{omgba2004numerical} for the simulations. The HEM is a set of nonlinear, hyperbolic one-dimensional PDEs:
\begin{align} \label{paper_4:eq:multi_phase_pipe_equations}
\begin{split} 
&\partial_t (A_g\rho_g) +  \partial_x (A_g\rho_g u_m)= - \alpha_g L(\rho_m) \delta(x-x_l), \\
&\partial_t (A_l\rho_l) +  \partial_x (A_l\rho_l u_m)= - \alpha_l L(\rho_m) \delta(x-x_l), \\
&\partial_t (A\rho_m u_m) +  \partial_x \left(\rho_m u_m^2 A + p(\rho_g)A\right) = -\frac{\rho u |u|}{2} f_f(\rho, u), 
\end{split}
\end{align}
with boundary conditions $\rho_g u_m(0,t) = (\rho_g u_m)_0$ and $\rho_l u_m(0,t) = (\rho_l u_m)_0$ at the pipe inlet on the left side, and $p(L,t) = p_L$ at the outlet on the right side. $f_f$ is the wall friction, $Re$ is the Reynolds number, and $L$ is the leak size. $\delta$ is the Dirac delta function, ensuring that the leak is only active at $x=x_l$. $A$ is the cross section area of the pipe, $A_g$ is the area occupied by gas and $A_l$ the area occupied by liquid. Hence, for $\alpha_g\in[0,1]$ and $\alpha_l\in[0,1]$ being the fraction of gas and liquid, respectively, we have:
\begin{align}
    A_g = \alpha_g A, \quad A_l = \alpha_l A, \quad A = A_g + A_l, \quad \alpha_g + \alpha_g =1.
\end{align}
$\rho_l$ is the liquid density assumed to be constant, $\rho_g$ is the gas density, and $\rho_m$ is the mixture density (not to be confused with the probability density functions, $\rho$),
\begin{align}
    \rho_m = \alpha_g \rho_g + \alpha_l \rho_l.
\end{align}
The wall friction, $f_f$, is given by,
\begin{align}
    f_f = 2 \left( \left(\frac{8}{Re}\right)^{12} + (a + b)^{-1.5}  \right)^{1/12}, \quad  a = \left(-2.457 \ln\left( \frac{7}{Re}\right)^{0.9} + 0.27 \frac{\varepsilon}{2r}  \right)^{16}, \quad b = \left(\frac{37530}{Re}\right)^{16},
\end{align}
the Reynolds number, $Re$, is given by
\begin{align}
    Re = \frac{2r\rho_m u_m}{\mu_m}, \quad \mu_m = \alpha_g \mu_g + \alpha_l \mu_l,
\end{align}
and the leak size, $L$, is given by,
\begin{align}
    L(\rho_m) = C_v\sqrt{\rho_m(p(\rho_g)-p_{\text{amb}})}.
\end{align}
For specific values of all constants in equation \eqref{paper_4:eq:multi_phase_pipe_equations}, see Table \ref{paper_4:tab:pipeflow_parameters}.

\begin{table}[ht]
    \small
    \centering
    \caption{Parameters for the multi phase pipe flow equations, \eqref{paper_4:eq:multi_phase_pipe_equations}. Note that the discharge coefficient and the leakage location have values denoted by intervals, as they are the parameters to determine.}
    \begin{tabular}{llll} 
    \toprule
    Physical quantity & Constant &  Value & Unit\\ 
    \midrule
     Pipe length & $L$ & 5000 & m \\
     Diameter & $d$ & 0.2 & m \\
     Cross-sectional area & $A$ & 0.0314 & $\text{m}^2$ \\
     Speed of sound in gas & $c$ & 308 & m/s \\
     Ambient pressure & $p_{\text{amb}}$ & $1.01325\cdot 10^5$ & Pa \\
     Reference pressure & $p_{\text{ref}}$ & $1\cdot 10^5$  &Pa \\
     Reference density (gas) & $\rho_{g}$ & 1.26 & kg/$\text{m}^3$  \\
     Reference density (liquid) & $\rho_{l}$ & 1003 & kg/$\text{m}^3$  \\
     Inflow velocity & $v_0$ & 4.0 & m/s \\
     Outflow pressure & $p_L$ & 1.0e6 & Pa \\
     Pipe roughness & $\varepsilon$ & $10^{-8}$ & m \\
     Fluid viscosity (gas) & $\mu_g$ & $1.8\cdot 10^{-5}$ & $\text{N}\cdot\text{s}/\text{m}^2$ \\
     Fluid viscosity (liquid) & $\mu_l$ & $1.516\cdot 10^{-5}$ & $\text{N}\cdot\text{s}/\text{m}^2$ \\
     Temperature & $T$ & 278 & Kelvin \\
     Discharge coefficient & $C_d$ & $\left[1, 2 \right]$ & m  \\
     Leakage location & $x_l$ & $[10,5990]$ & m \\ 
     \bottomrule
\end{tabular}
    \label{paper_4:tab:pipeflow_parameters}
\end{table}

We discretize the PDEs using the nodal discontinuous Galerkin method \cite{hesthaven2007nodal} with Legendre polynomials for the modal representation and Lagrange polynomials for the nodal degrees of freedom. We use the Lax-Friedrichs discretization for the numerical flux and BDF2 for time stepping. The nonlinear equations coming from the implicit time stepping are solved using Newton's method, where the resulting linear systems are solved via an LU factorization. Furthermore, the Jacobian matrix is only updated every 500 time-steps to speed up the computations. 

The true state and observations used for the data assimilation are simulated using 3000 elements and third-order polynomials. The states are then evaluated on regular grid of 512 points. 

The training data is simulated using 2000 elements and second-order polynomials. Thereafter, it is evaluated on a regular grid consisting of 512 grid points. The equations are solved with a time-step size of 0.01, for $t=[0,120]$ seconds. Hence, there are 12000 time steps. The surrogate model is trained to take steps 10 times larger than the high-fidelity model. 

For testing the D-LSPF, we compute a high-fidelity particle filter solution with 5000 particles as a baseline to measure if we estimate the distributional information accurately. The high-fidelity solver uses 700 elements and 3rd order polynomials. 

The available observations are located at 8 spatial locations: $x$=(489.24, 978.47, 1467.71, 1956.95, 2446.18, 2935.42, 3424.66, 3913.89, 4403.13). Only pressure is observed. The observations arrive with a time frequency of 4s, corresponding to every 400 time steps of the PDE model. To ensure consistent performance across various configurations, we compute the metrics over 8 different test trajectories with varying leak locations and sizes and take the average. We compute the state and parameter RRMSE against the true solution, the state AMRMSE against a HF particle filter solution, the state and parameter NLL against the HF particle filter solution, as well as the Wasserstein-1 distance of the posterior parameter distribution against the HF posterior.

We use this test case as an ablation study: we use the same particle filter setting for all approaches and only replace the chosen architectures. We compare the presented architectures with three alternatives: One where the transformer-based AE layers are replaced with convolutional ResNet layers. The dimensionality reduction in this network is performed through strided convolutions and the dimensionality expansion is performed using transposed convolutions. The second alternative is using the proposed ViT architecture but without the Wasserstein distance and consistency regularization in the latent space for the autoencoder. Lastly, we compare with a setup that still uses the proposed ViT AE with regularization, but replaces the transformer based time stepping with a neural ODE (NODE) \cite{chen2018neural}. In all cases, a latent dimension of 8 is chosen.

\paragraph{Remark} We also trained a Fourier Neural Operator (FNO) \cite{li2020fourier} as a surrogate model, but without success. In all attempts, the solutions exploded after a certain number of time steps. This may be due to the fact that there is a clear discontinuity which is located at the parameter, $x_l$. Fourier series may not be good approximators for such tasks. For this reason, FNO results are omitted from the chapter.

\begin{table*}[ht]
    \footnotesize
    \centering
    \caption{Computation times using a GPU for the D-LSPF applied to the multi phase leak localization test case. All timings are computed with 5000 particles. The high-fidelity solver makes use of 100 CPU cores and the neural network uses one GPU. "P-" refers to parameter estimation and "S-" refers to state estimation.} 
    \begin{tabular}{l|c|cccc}
    \toprule
    &   HF &   Reg-ViT-Trans &   NoReg-ViT-Trans &    Reg-Conv-Trans &   Reg-ViT-NODE \\
    \midrule
      P-RRMSE $\downarrow$ &   $5.2\cdot 10^{-1}$ &   $\mathbf{9.6\cdot 10^{-3}}$ &   $1.1\cdot 10^{-2}$  &   $7.4\cdot 10^{-1}$ &   $1.5\cdot 10^{-2}$\\
      P-Wasserstein-1 $\downarrow$ & - &    169.9 &   \textbf{168.4} &   875.7 &   172.7 \\
      S-RRMSE  $\downarrow$ &   $7.9\cdot 10^{-2}$ &   $\mathbf{2.5\cdot 10^{-2}}$ &   $ 2.9 \cdot 10^{-2}$ &   $8.3 \cdot 10^{-2}$ &   $3.1\cdot 10^{-2}$ \\
      S-AMRMSE $\downarrow$ & - &   $\mathbf{4.3\cdot 10^{-3}}$ &   $4.4\cdot 10^{-3}$ &   $4.5\cdot 10^{-3}$ &   $4.5\cdot 10^{-3}$ \\
      P-NLL $\downarrow$ & - & \textbf{5.09} & 5.70 & 6.01 & 12.54 \\
      S-NLL $\downarrow$ & - & 16.87 & \textbf{15.37} & 18.37 & 17.32   \\
       Time (GPU) $\downarrow$ &   - &    24.89s &    24.90s &   23.58s &   \textbf{6.52} \\  
       Speed-up (GPU) $\downarrow$ & - &   1807.95 &   1807.23 &   1908.40 &   \textbf{6901.84} \\  
       Time (CPU) $\downarrow$ &   45000s &    332.3s &    317.0s &   328.8s &   \textbf{45.9s} \\  
       Speed-up (CPU) $\downarrow$ & - &   135.4  &   142.0 &   136.9 &   \textbf{980.8} \\  
     \bottomrule
\end{tabular}
\label{paper_4:tab:multi_phase_results}
\end{table*}

The results are summarized in Table \ref{paper_4:tab:multi_phase_results}. Our chosen architecture demonstrates superior performance in most metrics. Note in particular that the results using the convolutional AE are significantly worse compared to the ViT AE, emphasizing the advantages of the transformer-based architecture across different combinations. Furthermore, the D-LSPF outperforms the high-fidelity particle filter in estimating the leak location and size as well as the state. This can be explained from the fact that the D-LSPF is trained on higher resolution trajectories. As this came at a higher cost only at  the training stage, it does not add to the computation time when running the particle filter.

\begin{figure}[ht]
     \centering 
     \begin{subfigure}[b]{0.42\textwidth}
         \centering
         \includegraphics[width=.85\textwidth]{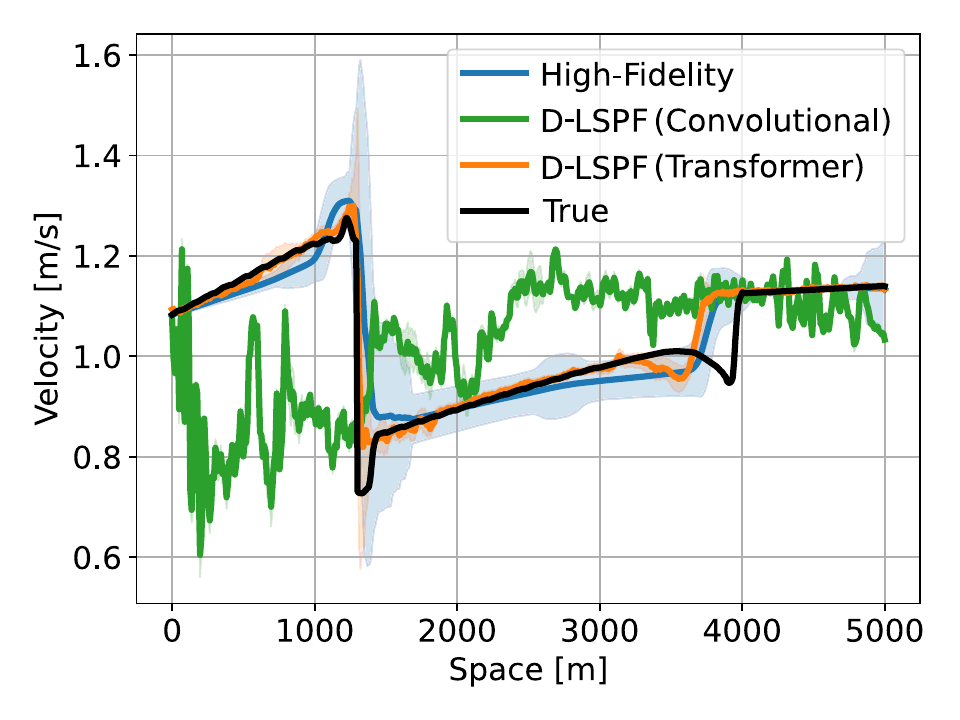}
         \caption{Velocity estimation at $t=40$s.}
         \label{paper_4:fig:multi_phase_state_short}
     \end{subfigure}
     \begin{subfigure}[b]{0.42\textwidth}
         \centering
         \includegraphics[width=.85\textwidth]{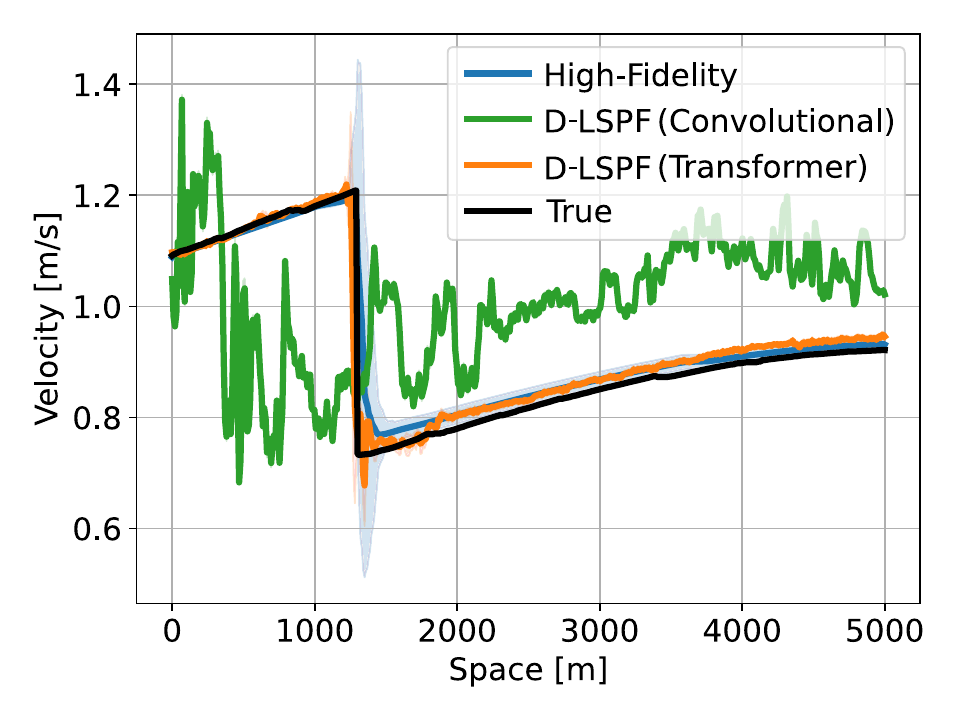}
         \caption{Velocity estimation at $t=120$s.}
         \label{paper_4:fig:multi_phase_state_long}
     \end{subfigure}
    \caption{Velocity estimation results for the multi phase pipeflow test case.}
    \label{paper_4:fig:multi_phase_state_results}
\end{figure}

In Figure \ref{paper_4:fig:multi_phase_state_results}, the true velocity at $t=40$ and $t=120$ are compared with estimates using a high-fidelity model, the D-LSPF with the ViT AE, and the D-LSPF with convolutional AE. The convolutional AE clearly fails to reconstruct the state in any meaningful way, while the D-LSPF with the ViT AE accurately estimates the velocity with the uncertainty concentrated around the leak location as expected. 

\begin{figure}
     \centering
      \begin{subfigure}[b]{0.42\textwidth}
         \centering
         \includegraphics[width=.85\textwidth]{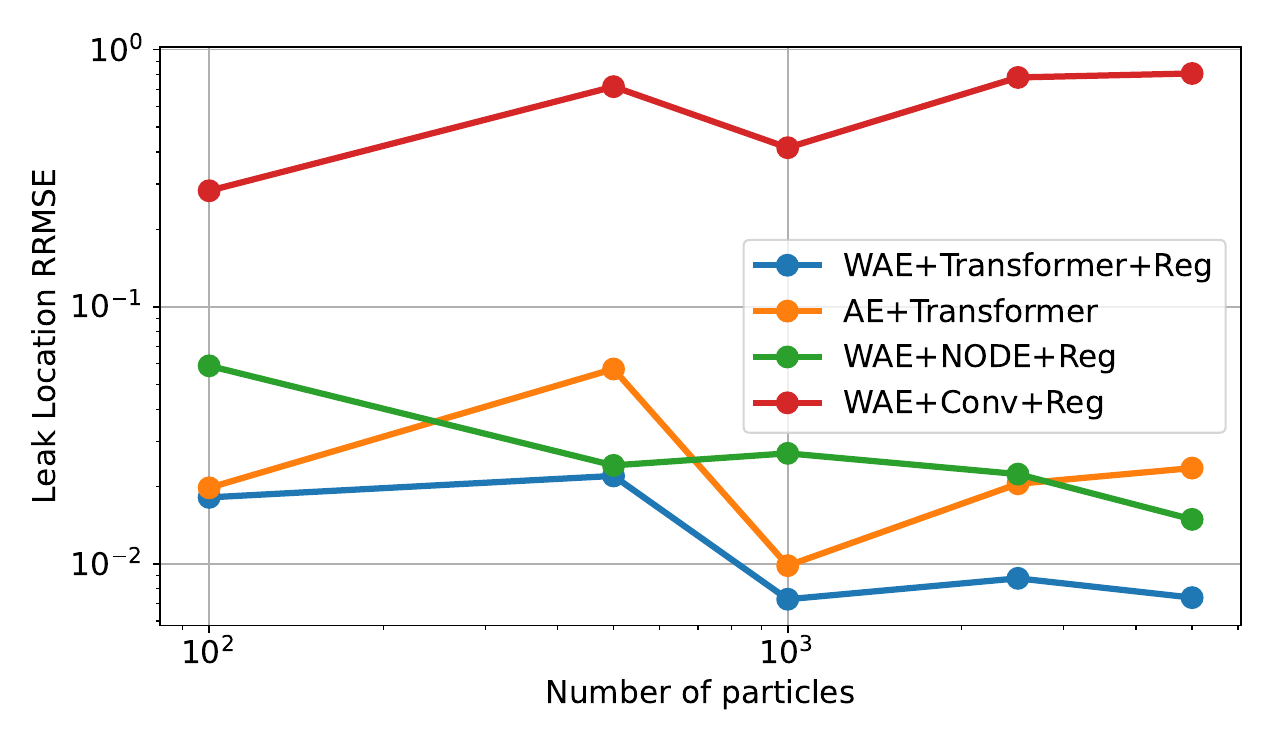}
         \caption{Leak location and size Relative RMSE}
         \label{paper_4:fig:multi_phase_pars_rrmse}
     \end{subfigure}
     \begin{subfigure}[b]{0.42\textwidth}
         \centering
         \includegraphics[width=.85\textwidth]{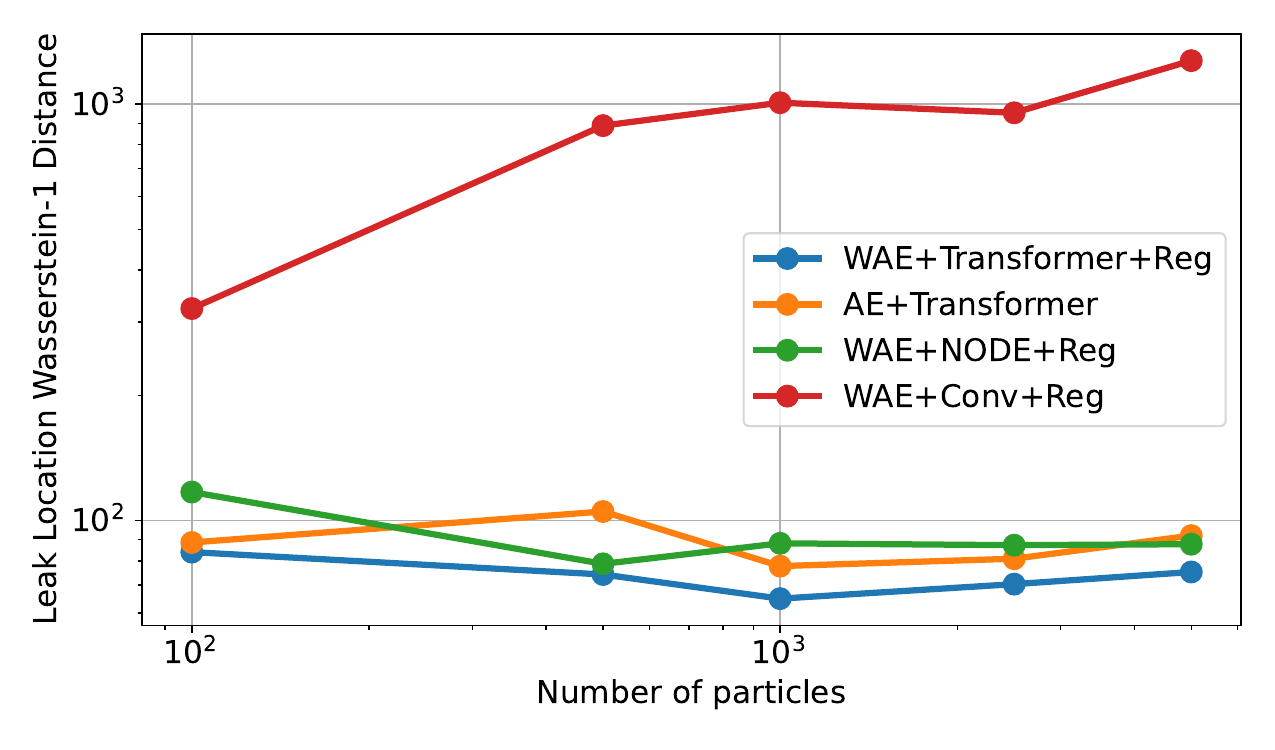}
         \caption{Wasserstein distance for the leak location. }
         \label{paper_4:fig:multi_phase_pars_wasserstein}
     \end{subfigure}
    \caption{Results for the leak size and location estimation at time 120s, for varying neural network specifications and number of particles.}
    \label{paper_4:fig:multi_phase_pars_results}
\end{figure}

In Figure \ref{paper_4:fig:multi_phase_pars_results}, we show the convergence accuracy of the leak location estimation for all the neural network setups. We see in Figure \ref{paper_4:fig:multi_phase_pars_rrmse} that the proposed setup is superior than the alternatives with respect to the RRMSE for all number of particles. Similarly, we reach the same conclusion with respect to the Wasserstein-1 distance of the posterior distribution of the leak location. 

\section*{Conclusion}\label{paper_4:section:conclusion}
We presented a novel particle filter, the D-LSPF, for fast and accurate data assimilation with uncertainty quantification. The D-LSPF was based on a surrogate model utilizing dimensionality reduction and latent space time stepping. The use of particle filters provided estimates for the state and parameters as well as for the associated uncertainties. For the AE, we made use of a novel extension of the vision transformer for dimensionality reduction and reconstruction. Furthermore, we discussed a number of regularization techniques to improve the performance of the D-LSPF. 

We demonstrated the D-LSPF on three different test cases with varying characteristics and complexities. In the first test, we compared with alternative deep learning-based data assimilation methods for the viscous Burgers equation. The D-LSPF showed superior performance by an order of magnitude regarding uncertainty estimation as well as almost 4 times better mean reconstruction. In the second test case, we performed state and parameter estimation on a wave tank experiment with few observations available in time and space. The D-LSPF performed up to 3 times better than alternative approaches while also being faster and successfully estimated both the state and parameter. Lastly, we applied the D-LSPF on a leak localization problem for multi-phase flow in a long pipe. We showed how the ViT AE and the regularization techniques drastically improved the state and parameter estimation: the D-LSPF provided speed-ups of 3 orders of magnitude compared with a high-fidelity particle filter while also being more accurate.

Several aspects of our work could benefit from future investigations. In particular, we made use of the bootstrap particle filter. While this particular version of the particle filter has many advantages such as relative ease of implementation and convergence, there are alternatives. It might be fruitful to analyze the quality of filters that utilize the differentiability of neural networks, such as the nudging particle filter with gradient nudging \cite{akyildiz2020nudging} or particle filters based on Stein variational gradient descent \cite{fan2021stein, maken2022stein}. 

In all, the D-LSPF significantly sped up complex data assimilation tasks without sacrificing accuracy, thus enabling fusing of increasingly complex models with data in real-time. This work promises a solid foundation for future digital twins.

\section*{Data availability}
The code for setting up and training the neural networks, including hyperparameter settings, for the test cases can be found in the GitHub repository \url{https://github.com/nmucke/latent-time-stepping}. The code for simulating training data for Burgers equations and the multi-phase leak location problem can be found in the same repository. The code for simulating the training data for the harmonic wave generation over a submerged bar test case can be made available upon request and in consultation with Associate professor Allan Peter Engsig-Karup. The code for the particle filter implementations as well as the test data can be found in the GitHub repository \url{https://github.com/nmucke/data-assimilation}. 

\bibliography{references}

\section*{Acknowledgements}
This work is supported by the Dutch National Science Foundation NWO under the grant number 629.002.213. The authors also acknowledge Oracle for providing compute credits for their cloud platform, Oracle Cloud Infrastructure. The authors furthermore acknowledge the help and code provided by Associate professor Allan Peter Engsig-Karup for the results related to the harmonic wave generation over a submerged bar test case.

\section*{Author contributions statement}
\textbf{N. M{\"u}cke:} Conceptualization, methodology, software, formal analysis, writing - original draft.  \textbf{S. Boht{\'e}:}  Funding acquisition, writing -review \& editing, supervision, project administration. \textbf{C. Oosterlee:} Funding acquisition, formal analysis, writing -review \& editing, supervision, project administration.

\section*{Competing interests}
The authors declare that they have no competing financial or personal interests that have influenced the work presented in this paper.

\end{document}